\documentclass[12pt,a4paper,twoside]{article}
\usepackage{epsfig}

\newcommand{\be}{\begin{equation}}
\newcommand{\ee}{\end{equation}}
\newcommand{\beqs}{\begin{eqnarray}}
\newcommand{\eeqs}{\end{eqnarray}}
\newcommand{\LL}{{\cal L}}

\newcommand{\tr}{{\rm tr}}
\newcommand{\Tr}{{\rm Tr}}
\newcommand{\half}{{1 \over 2}}
\newcommand{\Ad}{{A^\dagger}}
\newcommand{\ba}{{\bf a}}
\newcommand{\bb}{{\bf b}}
\newcommand{\bc}{{\bf c}}
\newcommand{\tha}{{\cal \theta}_1}
\newcommand{\thh}{{\cal \theta}_4}
\newcommand{\n}{{\rm n}}
\newcommand{\m}{{\rm m}}

\def\NP{{\it Nucl. Phys.\ }}
\def\PL{{\it Phys. Lett.\ }}

\def\PR{{\it Phys. Rev.\ }}
\def\PRL{{\it Phys. Rev. Lett.\ }}
\def\CMP{{\it Comm. Math. Phys.\ }}

\def\IJMP{{\it Int. Jour. Mod. Phys.\ }}
\def\MPL{{\it Mod. Phys. Lett. A\ }}

\expandafter\ifx\csname mathbbm\endcsname\relax

\else

\fi
\begin{document}
\begin{titlepage}
\begin{flushleft}  
       \hfill                      CCNY-HEP-06-9\\
       \hfill                       June 2006\\
\end{flushleft}
\vspace*{3mm}
\begin{center}
{\LARGE Physics and mathematics of Calogero particles \\}
\vspace*{8mm}
\large Alexios P. Polychronakos \\
\vspace*{5mm}
{\em Physics Department, City College of the CUNY \\
160 Convent Avenue, New York, NY 10031, USA \\
{\rm alexios@sci.ccny.cuny.edu}\/}\\

\vspace*{15mm}
\end{center}
%\maketitle

\begin{abstract}
We give a review of the mathematical and physical properties of the
celebrated family of Calogero-like models and related spin chains.

\end{abstract}

\end{titlepage}
\tableofcontents
\section{Introduction}

The system of one dimensional particles with inverse-square
pairwise interactions has assumed a ``Jack-in-the-box" role
in mathematical and theoretical physics over the last three
decades: it pops up in various seemingly disparate situations,
it is a recurring and unifying theme in high energy and condensed
matter physics and constitutes the prime example of an integrable
and solvable many-body system. Its range of applicability spans
fluid mechanics, spin chains, gauge theory and string theory.
This model has been labelled in the literature with 
various subsets and permutations of the last names of
Francesco Calogero, Bill Sutherland and Jurg Moser. We shall
simply call it the Calogero model, for brevity and in recognition
of its original inventor.

The Calogero model (and its many generalizations) has reached
maturity, in the sense that its various aspects (classical,
quantum, differential equation, operator, statistical, symmetry etc.)
have been exhaustively analyzed and no new groundbreaking
results seem to appear recently (I could be wrong!). It is unlikely,
however, that it will be relegated to the shelves of mathematical
physics for perpetuity. It has already enjoyed several revivals
in its 35-odd year history and there are still open issues awaiting
resolution. Chances are it has a few more surprises up its sleeve,
their unveiling requiring, and offering back, new intuition.

The purpose of this brief review is to expose the interesting
physical and mathematical properties of the system and whet the
appetite of readers for further study and research on the topic.
It is not meant to be exhaustive, or rigorous, or all-encompassing:
there are excellent extensive review articles that can serve this
purpose. The hope is that the small size, physical slant and stress
on newer developments will make this narrative an accessible and
motivational first piece.

\section{Basic features of the Calogero model}
\subsection{Introducing the model}

The prototype of the model is the system of identical particles scattering
on the line with inverse-square interaction potentials, as first
introduced by Calogero \cite{Calo}. Its hamiltonian (in an obvious
notation) is
\be
H = \sum_{i=1}^N \half p_i^2 + \sum_{i<j} \frac{g}{(x_i - x_j )^2}
\ee
The particle masses $m$ have been scaled to unity.

The motivation for such a model comes from various perspectives.
This is the only scale-free two-body potential that one can have:
quantum mechanically, the potential scales like the kinetic term and
therefore the coupling $g$ becomes dimensionless (in $\hbar =1$ units).
The inverse-square
potential would arise as a centrifugal term in higher dimensions
and is `borderline' from the sense of stability: anything stronger than
that at short distances and the particles would `fall' into each
other. Finally, from the long-distance point of view, the inverse
square potential is borderline for statistical mechanics phase
transitions: a stronger potential leads to phase transitions, while
a weaker one does not. The Calogero model `straddles the line' in
many respects.

To make the system bound, one has to introduce a kind of external
`box'. One way of doing this is to include an external
harmonic oscillator potential that confines the system. This is
nice, since it does not spoil the basic features of the system and
leads to an integrable model, the harmonic Calogero model.
Extracting thermodynamics from this system, however, is a bit tricky
since the box is not homogeneous and the process requires careful
scaling. Alternatively, one could put the system in a finite
periodic box (whose length can be scaled to $2\pi$). The particles
now interact through all the infinitely many periodic images of
themselves and the two-body potential becomes 
\be
V(x) = \sum_{n=-\infty}^\infty \frac{g}{(x+2\pi n)^2}
= \frac{g}{\left(2 \sin\frac{x}{2} \right)^2}
\ee
This is the Sutherland
model \cite{Suth}. There are other versions of this class of models
that we will not analyze. The original papers on the subject still
consist essential reading \cite{Calo}-\cite{PolEX}.
The classic report \cite{OP}
analyzes these systems in detail; the paper \cite{Rui} describes
relativistic analogs of these systems (known as the Ruijsenaars-Schneider
model) that are not touched in the present review; the lectures \cite{LesHouches}
cover many issues related to fractional statistics; and \cite{HP}
provides further mathematical results.

\subsection{Stability, hermiticity}

Classically the coupling constant $g$ should be positive to ensure
particles are not `sucked' into each other. Quantum mechanically
the uncertainty principle works in our favor and the minimum allowed
value for $g$ is $g= -\frac{1}{4}$ (we put henceforth $\hbar =1$).
This result can be derived by regularizing the singularity of the
potential at $x_i - x_j =0$ and taking the limit, as done, e.g.,
in Landau and Lifshitz's classic quantum mechanics book.

The hermiticity properties of the system for negative coupling constants
has received a lot of attention and there are ways to make sense of
values even less than $g= -\frac{1}{4}$. It is not our purpose to
analyze these here, since it seems that the applications that
arise naturally are the ones respecting the above condition.

For later convenience, it is useful to parametrize $g$ in the fashion
\be
g = \ell(\ell-1)
\ee
in which case the minimum value is naturally obtained for $\ell = \half$.
Note, however, that there are generically {\it two} values of $\ell$
that give the same $g$, namely $\ell$ and $\ell' = 1-\ell$. For 
$-\frac{1}{4} <g< 0$, in particular, they are both positive. It is
possible to argue that only the value $\ell \geq \frac{1}{2}$ is
relevant, although there are good reasons to retain both of them
for the case of negative $g$. The possible use of that will become
clear in the sequel. Note, further, that the above expression for $g$,
when we reintroduce $\hbar$, becomes $g = \ell (\ell - \hbar )$.
In the classical limit we have $g = \ell^2$, which can also be thought
as the limit $\ell \gg 1$.

\subsection{Properties of the classical motion}

The above system is classically integrable, which means that there are $N$
integrals of motion in convolution, that is, $N$ functions on
phase space with vanishing Poisson brackets:
\be
\{ I_n , I_m \} =0 ~,~~~ n,m=1 \dots N
\ee
For the scattering system (no external potential) $I_1$ is the total
momentum, $I_2$ is the total energy amd the higher $I_n$ are higher
polynomials in the momenta also involving the two-body potentials.

The integrals of motion derive from the so-called Lax matrix of the model.
Consider the $N \times N$ hermitian matrices
\be
L_{jk} = p_j \delta_{jk} + (1-\delta_{jk} ) \frac{i\ell}{x_{jk}}
~,~~~
A_{jk} = \ell \delta_{jk} \sum_{s=1}^N \frac{1}{x_{js}^2} +
\ell (\delta_{jk} -1) \frac{1}{x_{jk}^2} 
\ee
with $\ell^2 = g$ and $x_{jk} = x_j - x_k$. Then, upon use of the 
equations of motion of the
Calogero model, the evolution of the elements of the matrix $L$ is
\be
{\dot L} = i [L,A]
\ee
A pair of matrices satisfying this equation is called a Lax pair, with
$L$ the Lax matrix. What the above equation means is that the evolution
of $L$ is simply a unitary conjugation $L \to U L U^{-1}$ generated by the
(time-dependent) matrix $A$. So all the eigenvalues of $L$ are conserved
and the traces
\be
I_n = \tr L^n ~,~~~ n=1,\dots N
\ee
are constants of motion. Obviously $I_1 = \sum_i p_i$, while explicit
calculation and the use of the identity
\be
\sum_{i,j,k~{\rm distinct}} \frac{1}{x_{ij} x_{ik}} = 0
\ee
(easily proved by multiplying numerator and denominator by $x_{jk}$ and
cyclically redefining the dummy indices $i,j,k$) shows that $I_2 = 2 H$.
$I_3$ and beyond generate new, nontrivial conserved quantities.

It can further be shown that the above interals are in involution, that
is, they have vanishing Poisson brackets. For the scattering system it
is quite easy to give a physical proof: at $t \to \infty$ the particles
fly far away from each other and $L$ becomes $diag (p_i )$, the off-diagonal
elements asymptotically vanishing. The integrals of motion become simply
\be
I_n = \sum_i k_i^n
\ee
with $k_i$ the asymptotic momenta. Obviously these are in involution. 
Since the Poisson brackets of conserved quantities are also conserved, 
they must vanish at all times.

The key interesting property of the above model, which
sets it apart from other merely integrable models, is that, both 
classically and quantum mechanically, it mimics as closely as 
possible a system of free particles. We shall give here an overview
of these properties, without insisting on proofs, and will
come back to their derivation in subsequent sections.

Let us first look at its
classical behavior. The motion is a scattering event.
Asymptotically, at times $t=\pm \infty$,
the particles are far away, the potentials drop off to zero and 
motion is free. When they come together, of course, they interact and
steer away from their straight paths. Interestingly, however, when
they are done interacting, they resume their previous free paths
as if nothing happened. Not only are their asymptotic momenta the
same as before scattering, but also the asymptotic positions (scattering
parameters) are the same. There is no time delay of the particles
at the scattering region. The only effect is an overall reshuffling
of the particles. Thus, if one cannot tell particles from each other,
and if one only looks at scattering properties, the system looks free!

Similar behavior is exhibited by the harmonic Calogero and the Sutherland
model. The motion of the harmonic model is periodic, with period
determined by the harmonic oscillator potential. The particles seem to
revolve around ghostly paths of would-be non-interacting particles
in the external potential, their interaction creating deviations but
never completely upsetting the system. For the Sutherland model the
particles seem bound to a set of free particle paths corresponding to
a given set of momenta, deviating from them as they come near each other
and resuming them as they separate. The momenta of the free paths
(called `pseudomomenta')
are related to the integrals of motion of the system.

Is there a concise manifestation of the coupling constant $g$ in the
classical motion of the particle, except its details? It turns out
that there is a particularly neat result involving $g$:
the classical action of two particles scattering off each other is
the same as the action of free particles with the same asymptotic
momenta, but diminished by
\be
\Delta S = \pi \sqrt{g} = \pi \ell
\ee
Semiclassically,
the action corresponds to the phase of the wavefunction for the
scattering process. This suggests that the scattering phase shift
differs by a non-dynamical constant from the corresponding free
particle result. As we shall see, this is fully born out in
quantum mechanics.

\subsection{Properties of the quantum system: fractional statistics}

The above behavior carries over to quantum mechanics. The asymptotic
scattering momenta are the same before and after scattering. The 
classical fact that there is no time delay translates into the quantum
fact that the  scattering phase shift is independent of the momenta (remember
that the time delay is the momentum derivative of the phase shift).
Thus, it can only
be a function of the coupling constant and the total number of
particles. It is, actually, a very suggestive function:
\be
\theta_{sc} = \frac{N(N-1)}{2} \ell \pi
\ee
So the phase is simply $\ell \pi$ times the total number of
two-body exchanges that would occur in the scattering of free particles.
This squares with the previous classical result: the classical limit
corresponds to $\ell \gg 1$, so $\sqrt{g} \simeq \ell$ and we get
$\Delta S = \theta_{sc}$.

We can interpret the above property as the fact that Calogero particles
are essentially free but obey generalized statistics, as defined by their
scattering phase shift \cite{PolA}.
Clearly the case $\ell =0$ would correspond to free bosons and $\ell
=1$ to free fermions (for these two values the potential vanishes
and the system is, indeed, free). For any other value we can say that
the system has fractional statistics of order $\ell$. Note that, in
contrast to anyons, $\ell$ is {\it not} a phase: it can be fractional
but also bigger than one (``superfermions"?).

A word on the permutation properties of this system is in order.
The inverse-square potential is quantum mechanically impenetrable, and
thus the `ordinary' statistics of the particles (symmetry of the
wavefunction) is irrelevant: if the particle coordinates are in one of 
the $N!$
ordering sectors they will stay there for ever. The wavefunction 
could be extended to the other sectors in a symmetric, antisymmetric
or any other way, but this is irrelevant for physics. No interference
between the sectors will ever take place. All states have a trivial
$N!$ degeneracy. (That is to say all states in all irreps of $S_N$
have the same physical properties.) Permutation statistics are 
therefore irrelevant and we can safely talk about the effective 
statistics as produced by their coupling constant $\ell$.

Let us also review the properties of the confined systems. In the presence
of an external harmonic potential of the form
\be
V = \sum_i \half \omega^2 x_i^2
\ee
the energy spectrum of a system of uncoupled particles would be
\be
E = \frac{N}{2} \omega + \sum_i n_i \omega
\label{Efree}
\ee
The $n_i$ are nonnegative integers satisfying
\be
n_1 \le \dots \le n_N ~~~{\rm for~bosons}
\ee
\be
n_1 < \dots < n_N ~~~{\rm for~fermions}
\ee
The actual spectrum of this model is
\be
E = \frac{N}{2} \omega + \ell \frac{N(N-1)}{2} \omega + \sum_i n_i \omega
\ee
with $n_i$ being `excitation numbers' obeying bosonic selection rules:
$n_i \le n_{i+1}$.
Defining the `pseudo-excitation numbers'
\be
{\bar n}_i = n_i + (i-1) \ell
\ee
we can check that the expression of the spectrum in terms of the ${\bar n}_i$
is identical to the free one (\ref{Efree}) but with the quantum numbers
now obeying the selection rule
\be
{\bar n}_i \le {\bar n}_{i+1} - \ell
\label{nselect}
\ee
This is a sort of exclusion principle that requires the particle quantum
numbers to be at least a distance $\ell$ apart (as contrasted to 1 for
fermions and 0 for bosons). Again, a generalized statistics interpretation
is manifest \cite{Isa}. 

Let us clarify that the above numbers ${\bar n}_i$ are no more integers.
They do, however, increase in integer increments. The rule is that
the ground state is determined by the minimal allowed nonnegative
values for ${\bar n}_i$ obeying (\ref{nselect}) while the excited states
are obtained by all integer increments of these values that still
obey (\ref{nselect}).

The periodic (Sutherland) model has similar properties. Its spectrum is
\be
E = \sum_i \half k_i^2 + \ell \sum_{i<j} (k_j - k_i ) +
\ell^2 \frac{N(N^2 -1)}{24}
\ee
with the `momenta' $k_i$ being integers satisfying bosonic
rules: $k_i \le k_{i+1}$. This looks rather different than the corresponding
free expression (for $\ell=0$). Defining, however, again `pseudomomenta'
\be
p_i = k_i + \ell \left( i - \frac{N+1}{2} \right)
\ee
we can check that the expression for the spectrum becomes
\be
E = \sum_i \half p_i^2 
\ee
that is, the free expression. The pseudomomenta satisfy
\be
p_i \le p_{i+1} -\ell
\label{pselect}
\ee
that is, the same selection rule as the ${\bar n}_i$ before. Again,
we observe a generalization of the fermionic and bosonic selection
rules corresponding to statistics $\ell$. The ground state is the
minimal (nearest to zero) numbers satisfying (\ref{pselect}) while
excitations correspond to integer increments (or decrements) of
the $p_i$ still satisfying (\ref{pselect}).

Let us note that the above rule for $\ell =1$ reproduces the 
fermionic spectrum of particles with periodic boundary conditions
for odd $N$ and {\it anti}-periodic ones for even $N$: in the 
odd (even) $N$ case the momenta are quantized to (half-) integers.
This has a natural interpretation: when we take a particle around
the circle it goes over $N-1$ other particles. If we require the
phase shift of the wavefunction in this process to agree with the
minus signs picked up from the $N-1$ fermion exchanges we recover
the previous rule. We stress that, for free particles, this is
not a consistency requirement but rather an aesthetic rule. At any
rate, this is what the Sutherland model chooses to do!

In conclusion we see that the Calogero model can be though of as a
system of particles obeying generalized statistics. This manifests
in terms of the scattering phases and, most significantly for 
statistical mechanics, through a peculiar `level repulsion' of 
their quantum numbers generalizing the Fermi exclusion principle.

\subsection{Large-N properties of the Calogero model and duality}

Let us examine, now, the properties of the Calogero model as the
number of particles grows large. At zero temperature, a 
non-interacting
fermion system would form a Fermi sea. The corresponding
`Fermi surface' in one dimension degenerates to points. For
the system in an external harmonic potential there is just
one point corresponding to the highest excitation $n_F = N-1$.
For the free periodic system we would have two Fermi momenta at
$\pm p_F = \pm \frac{N-1}{2}$. Excitations over this ground state are,
then, conveniently classified in terms of particles (a filled Fermi
sea with an isolated particle above or below) and holes (a filled
sea with one unoccupied state inside it). 

Interestingly, the Calogero model presents a similar picture. The
qualitative features of both the Calogero and the periodic Sutherland model
are similar, so we pick the latter as most closely representing a
gas of free particles in a box. From (\ref{pselect}) above we
see that the ground state also forms a `pseudo-Fermi sea' (or should
we call it a `Luttinger sea'?) with Fermi levels rescaled by a
factor $\ell$: $p_F = \ell \frac{N-1}{2}$. Its minimal excitations
are analogous to the ones of a Fermi sea, but not quite:

$\bullet$ A particle would be an isolated occupied pseudomomentum
above or below a completely filled sea of pseudomomenta. Particles

\noindent
--are excited in units of 1 (the increments of their pseudomomentum).

\noindent
--take up a space $\ell$ in pseudomomentum (since they cannot be
`packed' closer than $\ell$ units apart).

$\bullet$ A hole would be an isolated empty space inside an
otherwise occupied sea. Interestingly, the minimal such excitation
is {\it not} obtained by removing one particle from the sea, but
rather by incrementing all pseudomomenta of the sea above the place
where we want to create the hole by one unit. Holes

\noindent
--are excited in units of $\ell$. Indeed, since the distance of
pseudomomenta in the sea is $\ell$, the possible positions of the hole
are at distances $\ell$ apart.

\noindent
--take up a unit space in pseudomomentum. Indeed, incrementing all
pseudomomenta above a given place in the sea by {\it two} units
creates two holes in that place, and so on; by locally reshuffling 
pseudomomenta we can then separate these holes.

Note that holes are {\it not} antiparticles. Removing a 
particle for the sea creates a gap of $\ell$ spaces and, from above,
$\ell$ holes. The correspondence is
\be
1~{\rm particle} \sim -\ell~{\rm holes}
\ee
We already observe a sort of duality between the two types of 
excitations. This can be summarized as
\be
{\rm particle} \leftrightarrow {\rm hole}~,~~~
\ell \leftrightarrow \frac{1}{\ell} ~,~~~
p \leftrightarrow \ell p
\ee
Under the above, the spectrum of excitations of the model remains
invariant. This is the simplest manifestation of a coupling-constant
duality that goes over to the correlation functions and Green's functions
of the model \cite{Gau,MPCF,LPS,Ha}. 
Obviously, this duality is spoiled by nonperturbative
effects, since holes are confined within the sea while there is no
`ceiling' for particle excitations.

\section{Particle symmetries and the Calogero model}

So far we have talked about the Calogero model and its properties
considering it a `given' system, with only circumstantial motivation.
It is now time to `derive' the model, in the sense of obtaining it by
starting from a set of principles.

The angle we will take is the one of the description of a set of
indistinguishable particles and their symmetries. Field theorists are
used to the idea of viewing particles as field quanta, or as representations
of the Poincar\'e group. The many-body (first quantized) point of view,
however, affords a perspective revealing the Calogero model as the
natural generalization of free indistinguishable particles.

The main idea is that {\it identical} particles admit the permutation group
as a dynamical symmetry commuting with the hamiltonian, while 
{\it indistinguishable} particles elevate the permutation
group to a (discrete) {\it gauge symmetry}. 
In plain words, configurations in which
particles are permuted correspong to the same physical state
and constitute `gauge' copies of the system.

There are two distinct ways to deal with a gauge system:

\noindent
$\bullet$ Reduce the system to a set of gauge invariant observables

\noindent
$\bullet$ Realize the gauge symmetry as a symmetry of the Hilbert space
and impose gauge constraints on the states

We shall explain below how each of these will lead to (versions of)
the Calogero model.

\subsection{Reducing to gauge invariant observables}

In the sense exposed above, phase space particle coordinates $x_i , p_i$
are not physical observables since they insist on assigning a label to each
particle; they are not permutation invariant. A set of invariants can be
constructed in terms of symmetric functions of the above coordinates.
Such a set is
\be
I_{n,m} = \sum_{i=1}^N : x_i^n p_i^m :
\ee
with $n,m \geq 0$ and $: \cdot :$ denoting a specific ordering. E.g., the
symmetric (Weyl) ordering between $x_i$ and $p_i$ can be adopted, which also
ensures the hermiticity of $I_{m,n}$.

The above invariants are, in general, overcomplete. Even if we were to
truncate the range of $n,m$ to $N$, there would still be of the order of
$N^2$ observables, while the number of independent basic operators is $2N$.
Classically, this reduncancy manifests in the existence of algebraic
identities between various $I_{m,n}$.

Quantum mechanically, overcompleteness translates into the presence of
Casimirs in the algebra of $I_{m,n}$. Indeed, the $I_{m,n}$ satisfy (a
particular parametrization of) the so-called $W_N$ algebra. In the
$N \to \infty$ limit their commutation relations can be conveniently 
repackaged into the `sine algebra' \cite{FZ}, by defining
\be
I(k,q) = \sum_{m,n=0}^\infty \frac{k^m q^n}{m! n!} I_{m,n}
\ee
with $k,q$ continuous `Fourier' variables. Then, assuming Weyl ordering
for the $I_{m,n}$, the $I(k,q)$ satisfy
\be
[ I(k,q) , I(k' , q' ) ] = 2i \sin \frac{kq' -k' q}{2} I(k+k' , q+q' )
\ee
from which the commutators of $I_{m,n}$ can be obtained by Taylor expanding
$I(k,q)$ and matching coefficients of $k^m q^n$. To lowest order in $\hbar$
we obtain
\be
[ I_{m,n} , I_{m' n' } ] = i (mn' - nm' ) I_{m+m' -1 , n+n' -1} + O(\hbar^2 )
\ee
which is the `classical' $W_\infty$ algebra. (We have put $\hbar =1$, but
lowest order in $\hbar$ corresponds to lowest order in $m,n$.)

For finite $N$ the algebra becomes nonlinear due to the presence of identities.
Alternatively, we can keep the full tower of $I_{m,n}$ and effectively impose
the identities as relations for the Casimirs of the algebra. Indeed, it is known
that the above algebra admits a host of representations, corresponding to the
underlying particles being bosons, fermions, parabosons or parafermions and
various other possibilities.

To see how the Calogero system emerges as one of these possibilities, concentrate,
for the moment, to the special case of two particles \cite{LM}. Their center of mass coordinate
and momentum
\be
X = \frac{x_1 + x_2}{2} ~,~~~ P = p_1 + p_2
\ee
are certainly gauge invariant observables; they correspond to $I_{1,0}$ and $I_{0,1}$.
The relative coordinate and momentum, however,
\be
x= x_1 - x_2 ~,~~~ p = \frac{p_1 - p_2}{2}
\ee
are not, since they are odd under permutation. We can form quadratic invariants as
\be
A= x^2 ~,~~~ B= \frac{xp+px}{2} ~,~~~ C= p^2
\label{ABC}
\ee
They correspond to
\be
A = 2 I_{2,0} - I_{1,0}^2 ~,~~~ B = I_{1,1} - \frac{1}{4} (I_{1,0} I_{0,1} + I_{0,1}
I_{1,0} ) ~,~~~ C = \frac{1}{2} I_{0,2} - \frac{1}{4} I_{0,1}^2
\ee
The relative variables $A,B,C$ commute with the center of mass variables and
close to the $SL(2,R)$ algebra
\be
[A,B] = i 2A ~,~~~ [B,C] = i 2B ~,~~~ [A,C] = i 4B
\ee
Classically, the above variables satisfy the constraint $AC=B^2$. Quantum mechanically,
this translates to the Casimir
\be
G = \frac{AC + CA}{2} - B^2
\ee
Physical Hilbert spaces correspond to irreducible representations of the
algebra of $A,B,C$, along with $X,P$.

The representation with vanishing
Casimir corresponds to the original system of two bosons or two fermions,
in which $A,B,C$ can be realized as in (\ref{ABC}). This is not, however,
the only one. A nonzero value for $G$, corresponding to a quantum correction
to the classical value, would also be a legitimate realization of the
indistinguishable particle dynamics. Interestingly, unitarity mandates
that $G \geq -\frac{1}{2}$, so we may parametrize
\be
G = \ell(\ell -1)
\ee
in direct analogy to the Calogero case. We observe that the representation
with the above value of $G$ can be realized as
\be
A = x^2 ~,~~~ B =\frac{xp+px}{2} ~,~~~ C= p^2 + \frac{\ell (\ell -1)}{x^2}
\ee
Effectively, the relative kinetic energy of the particles has acquired
an inverse-square potential part. The free particle hamiltonian for this
system would become
\be
H = \frac{P^2}{2} = \frac{1}{2} p_1^2 + \frac{1}{2} p_2^2 + 
\frac{\ell (\ell -1)}{(x_1 - x_2 )^2}
\ee
This is the Calogero model! The Calogero coupling plays the role of the
Casimir. The point is that the hamiltonian and
other observables of the Calogero model and the corresponding `free'
system (free of inverse-square interactions) in terms of physical
variables $X,P;A,B,C$ are identical.

For more than one particles the approach can be extended and we again
recover the Calogero model as one particular realization of the
indistinguishabe particle system.

The Hilbert space of the above realization, in terms of $x_i , p_i$, 
still provides a defining $N!$-dimensional realization of the permutation
algebra. For generic values of the Calogero coupling, however, this is
inconsequential: the inverse-square potential is impenetrable (both classically
and quantum mechanically) and thus the $N!$ sectors do not mix and are
physically equivalent. There is no need to decompose the defining
representation into irreducible components. The `normal' statistics of
the particles (wavefunction symmetry properties) have become immaterial,
being totally supplanted by the Calogero dynamics. Only for $G=0$, that is,
$\ell = 0$ or $1$, the Calogero
interaction vanishes and ordinary statistics come back into play. By
analytic continuation of the wavefunction properties, we can map bosons
to the $\ell =0$ case and fermions to the $\ell =1$ case.

\subsection{Augmenting the symmetry: Matrix Model}

The opposite way to realizing a gauge system is to keep the original
redundant formulation and impose gauge invariance as an operator
relation on the states (Gauss' law). In terms of the single-particle
phase space variables $x=i , p=i$ the gauge symmetry is the permutation
(symmetric) group $S=N$. This leads to the well-known and analyzed
cases of fermions, bosons and their parastatistics generalizations.

A different approach is to start with an {\it augmented} system, in
which both the dynamical variables and the gauge symmetry have been
expanded, giving at the end the same gauge invariant degrees of freedom
\cite{GMM}.
Specifically, we could formulate the particles in terms of
the eigenvalues of an $N \times N$ matrix. There is no a priori
ordering of these eigenvalues, so this certainly encodes identical
particles. It is clear that the permutation symmetry of the
problem has been promoted to the continuous symmetry of unitary
conjugations of this matrix, which leaves the eigenvalues intact.

This unitary conjugation symmetry is now the gauge group. States 
need not necessarily
be singlets under this symmetry, however, just as in the previous
section the Casimir $G$ needed not vanish. We can simply select the
Hilbert space to transform under an irreducible representation
of the gauge group and identify states transforming to each other
under the action of the group as a unique physical state, in direct
analogy to the considerations that lead to parastatistics.

This is the celebrated matrix model formulation, to be fully analyzed
in the subsequent sections. It has found various applications in
physics, the most directly related to identical particles, perhaps,
being the noncommutative Chern-Simons description of the finite
quantum Hall droplet \cite{APQH}-\cite{HKKCR}.

\section{The hermitian matrix model}
\subsection{Classical analysis}

Let us first examine a
matrix model that parallels as closely as possible particle mechanics
on the infinite line.
The kinematical variable is a hermitian $N \times N$ matrix $M$ and the
lagrangian reads 
\be
\LL = \tr \left\{ \half {\dot M}^2 - V(M) \right\}
\ee
$V(x)$ is a scalar potential evaluated for the matrix variable $M$.

Clearly the above has a time-translation invariance which leads to the
conserved energy
\be
H = \tr \left\{ \half {\dot M}^2 + V(M) \right\}
\ee
Moreover, the action is invariant under time-independent unitary 
conjugations of the matrix $M$:
\be
M \to U M U^{-1}
\ee
This nonabelian $SU(N)$ symmetry leads to the conserved  hermitian
traceless matrix
\be
J = i [M, {\dot M} ]
\ee
where $[ ~,~ ]$ denotes ordinary matrix commutator. These are the
`gauge charges' that, when fixed, will determine the particular realization
(`statistics') of the indistinguishable particle system.
But let us further analyze the implications of fixing these
charges classically.

We are interested in the dynamics of the eigenvalue of $M$, so we
parametrize it as
\be
M = U \Lambda U^{-1}
\ee
where $U(t)$ is the unitary `angular' part of the matrix and $\Lambda (t)
= diag\{x_1 , \dots x_N \}$ are the eigenvalues. Clearly the conserved
quantity $J$ has to do with invariance under `rotations' of the angular
part of $M$ and thus corresponds to the `angular momentum' of $U(t)$.
We define the `gauge potential'
\be
A = - U^{-1} {\dot U}
\ee
$\dot M$, $J$ and the lagrangian $\LL$ become, in this parametrization,
\begin{eqnarray}
{\dot M} &=& U \left( {\dot \Lambda} + [ \Lambda , A ] \right) U^{-1}
\label{dotM} \\
J &=& i U \left( \left[ \Lambda, [ \Lambda , A ]
\right] \right) U^{-1} \equiv U K U^{-1} \\
\LL &=& \tr \left\{ \half {\dot \Lambda}^2 + [ \Lambda , A ]^2
- V( \Lambda ) \right\} \\
&=& \half \sum_{i=1}^N {\dot x}_i - \half \sum_{i,j=1}^N ( x_i - x_j )^2
A_{ij} A_{ji}
\label{LLU}
\end{eqnarray}
The matrix elements of $A$ and $K$ are related
\be
K_{jk} = i \left[ \Lambda, [ \Lambda , A ] \right]_{jk}
=  i ( x_j - x_k )^2 A_{jk}
\label{Kjk}
\ee
Finally, solving (\ref{Kjk}) for $A_{jk}$ and putting into (\ref{LLU}) 
we obtain
\be
\LL = \sum_i \half {\dot x}_i^2 + \half \sum_{i \neq j} \frac{
K_{ij} K_{ji} }{( x_i - x_j )^2} -\sum_i V( x_i )
\ee
The first two terms are kinetic, coming from ${\dot M}^2$, while the
last one is potential. Therefore, the hamiltonian $H$ is
\be
H = \sum_i \half p_i^2 + \half \sum_{i \neq j} \frac{
K_{ij} K_{ji} }{( x_i - x_j )^2} +\sum_i V( x_i )
\label{HxK}
\ee
Note that the eigenvalues are kinematically coupled by an inverse-square
type potential with the angular momentum degrees of freedom. The connection
of the matrix model to the Calogero model along the lines presented
here and below was first established in \cite{KKS}. Also, the hamiltonian
(\ref{HxK}) has been proposed independently of the matrix model
as an $SU(N)$-generalization of the classical Calogero system
\cite{GHW}.

We can now examine special cases:

a) The most `gauge invariant' sector is,
of course, the one in which the angular momentum charges vanish, that is,
$J=0$. In that case, (\ref{HxK}) for $K=0$ becomes the hamiltonian of
non-interacting particles in an external potential $V(x)$. This would
be the case of `standard' particles.

b) For the next simplest case the angular momentum $J$ should be
as simple as possible without vanishing. Only its eigenvalues are
really relevant, since we can always perform a time-independent unitary
transformation $V$ which would shift $U \to VU$ and would rotate
$J \to V J V^{-1}$. The simplest choice would be to take the eigenvalues of 
$J$ to be equal. Unfortunately, this is not possible since the traceless
condition would make them vanish. The simplest possible choice is to take
all the eigenvalues equal to $\ell$ except one, which would cancel the
trace by being $(1-N) \ell$. This can be written in terms of an
arbitrary column $N$-vector $v$ as
\be
J = \ell ( v v^\dagger -1) ~,~~~ v^\dagger v = N
\ee
in which case $K$ becomes
\be
K = \ell ( u u^\dagger -1) ~,~~~ u = U^{-1} v
\ee
{}From (\ref{Kjk}) we see that $K_{ii} = 0$ (no sum on $i$) and thus
\be
u_i u_i^* =1  ~~~ {\rm (no~sum)}
\ee
So the coefficient of the inverse-square potential in (\ref{HxK})
becomes
\be
K_{ij} K_{ji} = \ell u_i u_j^* ~ \ell u_j u_i^* = \ell^2 ~~~(i \neq j)
\ee
Finally, (\ref{HxK}) becomes
\be
H = \sum_i \half p_i + \sum_{i < j} \frac{
\ell^2 }{( x_i - x_j )^2} +\sum_i V( x_i )
\label{Hxell}
\ee
This is the Calogero model! The potential strength $g=\ell^2$ is
related to the conserved charge $\ell$. Quantum mechanically,
picking this charge will amount to a choice of statistics. We also get, 
at this stage, an arbitrary external potential $V(x)$.

c) More general choices of $J$ amount to more variety in its eigenvalues.
$K_{ij} K_{ji}$ now, in general, becomes time-dependent and the dynamics
more complicated. We postpone the discussion for the quantum case where
it will be shown that this corresponds to Calogero particles having also
internal degrees of freedom. This will be a generalization of the
discussion of the first section, with irreps of $SU(N)$ substituting
the irreps of $S_N$.

Now that we have this new approach we can use matrix technology to
demonstrate the integrability of the Calogero model \cite{KKS,OP}. 
For $V(x)=0$ the matrix motion becomes free and $\dot M$ is conserved.
The conjugation-invariant quantities
\be
I_n = \tr {\dot M}^n
\ee
are also conserved and in involution (the matrix elements of $\dot M$
are momenta and have vanishing Poisson brackets). From (\ref{dotM})
and (\ref{Kjk}) we have
\begin{eqnarray}
(U {\dot M} U^{-1} )_{jk} &=& \delta_{jk} {\dot x}_j - 
(1-\delta_{jk} ) \frac{ i K_{jk}}{ x_j - x_k } \\
&=& \delta_{jk} \, {\dot x}_j - 
(1-\delta_{jk} ) \frac{ i u_j u_k^* }{ x_j - x_k } 
\end{eqnarray}
(note the similarity of the above expression for $U {\dot M} U^{-1}$ 
with the Lax matrix defind in section 2.3).
When the above expression is inserted in the trace $I_n = \tr {\dot M}^n$
clearly $U$ drops and products of the form $u_i u_j^* \, u_j u_k^*
\dots u_i^*$ will appear which reduce to powers of $\ell$. Therefore,
the $I_n$ reduce to expressions involving only $x_i$, ${\dot x}_i$
and the coupling constant $\ell$. These are the conserved integrals of
the Calogero model.

Starting from the matrix model the actual motion of the Calogero
model can be obtained. The solution for $M$ is
\be
M = B+Ct
\ee
for arbitrary matrices $B,C$. The conserved charge becomes
\be
J = i [M, {\dot M} ] = i [B,C] = i \ell (u u^\dagger -1)
\label{JBC}
\ee
By unitary transformations we can choose the phases of $u$ such
that $u_i =1$; choices for $B,C$, then, that satisfy (\ref{JBC}) are
\be
B_{jk} =  \delta_{jk} \, q_j ~,~~~
C_{jk} = \delta_{jk} p_j - (1-\delta_{jk} ) \frac{ i \ell }
{ q_j - q_k } 
\ee
$q_i$ and $p_i$ are the initial conditions for $x_i$ and
${\dot x}_i$ at time $t=0$. Diagonalizing, then, $M=B+Ct$ for the
above $B,C$ produces the motion of the system. Another choice,
in which $C$ is diagonal, is
\be
B_{jk} = \delta_{jk} a_j + (1-\delta_{jk} ) \frac{ i \ell }
{ k_j - k_k } ~,~~~ C_{jk} = \delta_{jk} k_j
\ee
$k_i$ and $a_i$ are asymptotic momenta and impact parameters.
For $t \to \pm \infty$ the off-diagonal elements of $B$ produce a 
perturbation of order $t^{-1}$ to the eigenvalues, so the motion is 
determined by the diagonal elements $a_i + k_i t$ alone. We recover the
result that the motion at asymptotic regions is the same as if the
particles were free.

We conclude by proving that the matrix model is also integrable and
solvable in the presence of a harmonic oscillator potential
$V(x) = \half \omega^2 x^2$. The non-hermitian matrix $Q={\dot M}
+ i\omega M$ evolves as
\be
Q(t) = e^{i\omega t} Q(0)
\ee
and the matrix $Q^\dagger Q$ is conserved. We leave it as an exercise
to derive the conserved integrals and the motion of the corresponding
Calogero problem.

External potentials with up to quartic dependence on $x$ also lead to
integrable, although not so solvable, models \cite{IP}. 
It is an open question
to prove this is all there is, or to find yet more integrable potentials.

Finally, we may wonder what restricts us to one dimension. We chose
a model with one matrix, and its eigenvalues corresponded to coordinates
of particles on the line. We could, indeed, start with an appropriate model
with many matrices, which would reproduce particle motion in higher
dimensions \cite{PolC}. The integrability and solvability properties of 
such extended models, however, are much less pleasant. The question of
whether they represent a workable extension of identical particles remains open.

\subsection{The hermitian matrix model: quantum}

We will, now, perform the quantization of this system. We will do it
first for the hermitian matrix model and subsequently for the unitary one.
Each model has its own advantages and appears in different situations.

Consider the hermitian model with a quadratic potential,
\be
\LL = \tr \left( \half {\dot M}^2 - \omega^2 M^2 \right)
\ee
with $\omega$ a scalar frequency. The above can also be
written in terms of matrix elements
\be
\LL = \sum_{jk} \half |{\dot M}_{jk} |^2 - \omega^2 |M_{jk} |^2
\ee
This is nothing but $N^2$ harmonic oscillators. So the system is
in principle trivial and solvable. All its nontrivial features
emerge from the reduction to a subspace corresponding to a fixed
value for the `angular momentum' $J$. Reducing $M$ to its matrix elements
is not beneficial for this purpose; we need to treat it as a matrix.

We begin by defining a canonical momentum matrix conjugate to
the `coordinate' $M$
\be
P = \frac{\partial \LL}{\partial {\dot M}} = {\dot M}
\ee
In terms of $M$ and $P$ the hamiltonian of the model becomes
\be
H = \tr \left( \half P^2 + \half \omega^2 M^2 \right)
\ee
The Poisson brackets are
\be
\{ M_{jk} , P_{lm}\} = \delta_{jm} \delta_{lk}
\ee

Upon quantization, the matrix elements of $M$ and $P$ become
operators and the above Poisson brackets become quantum mechanical
commutators (not to be confused with matrix commutators). The commutator
or $M_{jk}$ and $P_{lm}$ can be written conveniently by thinking of indices
$j,k$ as acting on a linear space $1$ and indices $l,m$ as acting
on linear space $2$. The 4-index symbol 
\be
\delta_{jm} \delta_{lk} \equiv (T_{12})_{jk;lm}
\ee
acts on both spaces and is, in fact, the operator exchanging the two
spaces. Denoting with $X_1$ and $X_2$ any matrix $X$ acting on space $1$
or space $2$, respectively, $T_{12}$ satisfies
\be
X_1 T_{12} = T_{12} X_2 ~,~~~ X_2 T_{12} = T_{12} X_1
\label{Tperm}
\ee
We also note the partial trace and unimodularity properties
\be
\tr_1 T_{12} = I_2 ~,~~~ \tr_2 T_{12} = I_1 ~,~~~ T_{12}^2 = I_{12}
\label{Ttr}
\ee
with $I_1$, $I_2$ and $I_{12}$ denoting the identity matrix in space $1$, space
$2$, or space $1 \times 2$ respectively. With the above notation the quantum
commutator of $M$ and $P$ becomes
\be
[ M_1 , P_2 ] = i T_{12}
\ee

The complex matrix $Q = {\dot M} + i \omega M$ introduced before, 
and its conjugate $Q^\dagger$,
are the matrix analogs of creation and annihilation operators. 
Defining matrix operators with the standard quantum normalization
\be
A^\dagger = \frac{1}{\sqrt{2\omega}} (P + i \omega M) ~,~~~
A = \frac{1}{\sqrt{2\omega}} (P - i \omega M)
\ee
they satisfy
\be
[ A_1 , A^\dagger_2 ] = T_{12} ~,~~~ 
[ A_1 , A_2 ] = [ A_1^\dagger , A_2^\dagger ] = 0
\label{comQ}
\ee

By analogy with the standard harmonic oscillator, let us define the matrices
\be
L = A^\dagger A ~,~~~~ R' = -A A^\dagger
\ee
The matrix operator $L = A^\dagger A$ is normal ordered, since all (quantum) creation
operators are to the left of annihilation operators. $R'$, on the other
hand, is not. We can define a new (quantum) normal ordered operator $R$, in which the
matrix multiplication is first performed in the order $A A^\dagger$ 
and {\it then} the creation matrix elements are moved to the left. Specifically
\be
R_{jk} = - : A_{js} (A^\dagger)_{sk} : = (A^\dagger)_{sk} A_{js}
\ee
Using the commutation relations of $A$ and $A^\dagger$ we see that $R$ and
$R'$ are trivially related:
\be
R = R' + N
\ee
In terms of $L$ and $R$, the hamiltonian can be written
\be
H = \frac{1}{2} \omega \tr (A A^\dagger + A^\dagger A) 
= \omega \tr L + \frac{N^2}{2} \omega =
-\omega \tr R + \frac{N^2}{2} \omega
\ee
From the basic commutators (\ref{comQ}) we can infer the commutation
relations:
\be
[L_1 , A_2^\dagger ] = T_{12} A_2^\dagger ~,~~~ 
[R_1 , A_2^\dagger ] = A_2^\dagger T_{12}
\ee
as well as their hermitian conjugates.
Tracing the first equation above with respect to space $1$, we obtain
\be
[ H , A^\dagger ] = \omega A^\dagger
\ee
This means that any matrix element of $A$ is a creation operator, creating
one quantum of energy $\omega$. This is hardly surprising, since the
matrix model is, indeed, $N^2$ harmonic oscillators with identical
frequency $\omega$.

The commutators of the matrix elements of $L$ and $R$ are calculated to be
\beqs
[ L_1 , L_2 ] &=& ( L_1 - L_2 ) T_{12} \cr
[ R_1 , R_2 ] &=& ( R_1 - R_2 ) T_{12} \label{LRcomm} \cr
[ L_1 , R_2 ] &=& 0
\eeqs
The above is nothing but two commuting copies of the $U(N)$ algebra
in disguise. To see this, define the fundamental
$U(N)$ generators $T^a$, $a=0,1, \dots N^2 -1$, with $T^0 = I /\sqrt{N}$
satisfying the normalization condition
\be
\tr ( T^a T^b ) = \delta^{ab}
\ee
and the $U(N)$ commutation relations
\be
[ T^a , T^b ]_{\rm{matrix}} = i f^{abc} T^c
\ee
where $[ . , . ]_{\rm{matrix}}$ is a matrix commutator. Then expand $L$ and $R$ 
in the complete basis $T^A$ of hermintian $N \times N$ matrices:
\be
L^a = \tr (T^a L) ~,~~~ L= \sum_a T^a L^a
\ee
and similarly for $R^a$. The (scalar) expansion coefficients $L^a$, $R^a$,
upon use of (\ref{LRcomm}), satisfy
\beqs
[ L^a , L^b ] &=& i f^{abc} L^c \cr
[ R^a , R^b ] &=& i f^{abc} R^c \cr
[ L^a , R^b ] &=& 0
\eeqs
Since $L$ and $R$ mutually commute, their powers and traces commute
as well, so
\be
[ \tr L^n , \tr R^m ] = 0
\ee
On the other hand, using the fundamental commutation relation for
$A$ and $A^\dagger$ (\ref{comQ}) and the properties of $T_{12}$ 
(\ref{Tperm},\ref{Ttr}), we can relate the traces of $L$ and $R$ as
\beqs
\tr_1 L_1^n &=& \tr_1 (A_1^\dagger A_1 )^n 
~=~ \tr_1 \left[ A_1^\dagger \tr_2 T_{12} A_1 (A_1^\dagger A_1 )^{n-1} \right] \cr
&=& \tr_1 \tr_2 \left[ A_1^\dagger T_{12} A_1 (A_1^\dagger A_1 )^{n-1} \right]
~=~ \tr_{12} \left[ T_{12} A_2^\dagger A_1 (A_1^\dagger A_1 )^{n-1} ) \right] \cr
&=& -\sum_{k=0}^{n-1} \tr_{12} \left[ T_{12} (A_1 A_1^\dagger )^k T_{12} 
(A_1^\dagger A_1 )^{n-1-k} \right] + \tr_{12} \left[ T_{12} 
(A_1 A_1^\dagger )^{n-1} A_1 A_2^\dagger \right] \cr
&=& -\sum_{k=0}^{n-1} \tr ( A A^\dagger )^k \tr (A^\dagger A )^{n-1-k} + 
\tr (A A^\dagger )^n \cr
&=& \sum_{k=1}^n (-1)^k \tr (R-N)^{k-1} \tr L^{n-k} + (-1)^n \tr (R-N)^n
\eeqs
Working recursively with the above relation, we can express traces of $L$
entirely in terms of traces of $R$ and vice
versa; mutual commutativity of the two sets, therefore, translates into
commutativity of the elements of each set, which shows that the quantities
\be
I_n = \tr (A A^\dagger )^n
\ee
are conserved and in involution.

The above proof, involving two commuting matrix operators $L$ and $M$
that are subsequently related, may seem a bit `too slick'.
Nevertheless, the involution of
$I_n$ should be clear from (\ref{LRcomm}): the $I_n$ are nothing but the
Casimirs of the $U(N)$ algebra generated by $L^a$. $I_1 = \tr L \sim H$
is simply the $U(1)$ charge, while the rest are (related to) the higher
$SU(N)$ Casimirs. Since $SU(N)$ is of rank $N-1$ and has $N-1$ independent
Casimirs, we recover $N$ commuting charges altogether. The higher charges
$I_n$, $n>N$, are related to the fundamental charges $I_1 , \dots I_N$
via nontrivial, nonlinear, $N$-dependent relations.

\subsection{The hermitian matrix model: reduction and spectrum}

It remains to do the reduction to the `gauge' sectors, that is, fix
the charge $J = i [ M , \dot M ]_{\rm{matrix}}$ and work out its implications
for the quantum states of the system.

The `angular momentum' $J$ classically generates unitary conjugation of
the matrix $M$; this implies that quantum mechanically it will become an
$SU(N)$ algebra. This can be seen explicitly: from its definition $J$
can be expressed in terms of the operators defined in the previous section
\be
J = i :[M,P]_{\rm{matrix}}: \, = \, : [ A^\dagger , A ]_{\rm{matrix}} : 
\, = \, A^\dagger A \, - : A A^\dagger : \, = L+R
\ee
$L$ and $R$ are two commuting $U(N)$ algebras, therefore their sum is
another $U(N)$ algebra.

Classically $J$ is traceless and therefore its $U(1)$ part vanishes,
making it an $SU(N)$ matrix. Quantum mechanically this should still be
true, since it generates the transformation
\be
M \to U M U^{-1}
\ee
which has trivial $U(1)$ part. This is ensured by the normal ordering
of the above expreession for $J$. $L$ and $R$ satisfy $\tr L = - \tr R$
and so $\tr J = 0$.

Reduction of the system to particular values of $J$ corresponds, quantum
mechanically, to fixing the representation of the $SU(N)$ algebra $J$.
So the system decomposes into sectors labelled by the allowed irreducible
representations (irreps) of $J$. Further, states within each sector related
via the action of $J$ are identified as a unique physical state, since $J$
is a `gauge' symmetry.

To identify these sectors we first need to identify the possible irreps
for $L$ and $R$. This can be done by examining their form. Take $L$, at first:
it is nothing but $N$ copies of the Jordan-Wigner bosonic oscillator construction
of the $U(N)$ algebra over the fundamental representation. This may require
some explaining.

Let $R_{\alpha \beta}^a$ be any $d$-dimensional representation of a Lie group, where
$\alpha , \beta =1,\dots d$ label its
matrix elements and $a$ labels its generators. Define a set of $d$
commuting creation and annihilation operators $a_\alpha , a_\alpha^\dagger$ 
\be
[ a_\alpha , a_\beta^\dagger ] = \delta_{\alpha \beta}
\ee
The operator obtained by `sandwiching' the matrix $R^A$ between the
vectors $a^\dagger$ and $a$
\be
G^a = a_\alpha^\dagger R_{\alpha \beta}^a a_\beta
\ee
satisfies the commutation relations of the Lie algebra. Therefore, it
provides representations of the algebra, imbedded in the Fock space
of the oscillators. Specifically, it provides all representations generated
by the {\it fully symmetrized} direct product of any number of representations
$R$. This includes the singlet (the Fock ground state), $R$ itself (the set
of $d$ states with excitation number one) etc.

For the specific case of the $U(N)$ Lie algebra with $R$ the fundamental 
representation $F$, we need $N$ oscillators and we get
\be
G^a = a_j^\dagger T_{jk}^a a_k
\ee
Transforming $G$ from the generator basis $G^a$ to the matrix basis
$G = \sum_a G^a T^a$, $G$ simply becomes
\be
G_{jk} = a_j^\dagger a_k
\ee

In view of the above, the matrix elements of $L$ can be written
\be
L_{jk} = (A^\dagger )_{js} A_{sk} = A_{sj}^\dagger A_{sk}
\ee
It is clear that $j$ and $k$ play the role of the fundamental indices
of $U(N)$, while $s$ is a summation index that runs over $N$ values;
for each fixed $s$, $A_{sj}^\dagger$ and $A_{sk}$ play the role of
$a_j^\dagger$ and $a_k$, respectively, while for different $s$ all 
operators commute. So the above $L$ is the direct sum of
$N$ independent (commuting) Jordan-Wigner realizations of $U(N)$ over
the fundamental.

For each fixed $s$ this realization includes, as explained earlier,
all the totally symmetrix tensor products of the fundamental representation
of $U(N)$; that is, all irreps with a single row in their Young tableau.
The direct sum of $N$ such representations, however, includes irreps with
up to $N$ rows, which is the general case. We conclude that the spectrum of 
$L$ spans the {\it full set} of irreps of $U(N)$. The $U(1)$ charge ($\tr L$) is
simply the total excitation number and is given by the number of boxes of the irrep.

A Jordan-Wigner construction based on the antifundamental representation
${\bar T}^a = -({T^a})^* = -({T^a})^t$ would lead to an expression in terms 
of $N$ oscillators
\be
G_{jk} = - a_k^\dagger a_j
\ee
Comparing with the expression for $R$
\be
R_{jk} = - A_{ks}^\dagger A_{js}
\ee
we conclude that $R$ is the direct sum of $N$ independent Jordan-Wigner
realizations of $U(N)$ over the antifundamental, which again spans the full
set of irreps of $SU(N)$. The $U(1)$ charge, now, is the negative of the
total excitation number.

Representations of $L$ and $R$ are, in fact, constrained to be
conjugate to each other. This arises due to their construction
in terms of the same bosonic creation and annihilation operators,
and is also manifest by the relation between their traces, which makes
all their even (odd) Casimirs equal to (minus) each other. So the representations
that $J=L+R$ can carry are of the form $r \times {\bar r}$. Such representations
include the singlet, the adjoint etc. and they have as common property that
their $Z_N$ charge vanishes. We conclude that the irreps of $J$ must have
a number of boxes in their Young tableau that is {\it an integer multiple
of $N$}.

The above result for $J$ could have been obtained by noticing that $J$ itself
is a bosonic Jordan-Wigner construction over the {\it adjoint} representation
of $SU(N)$. Clearly the adjoint and all its symmetric products have a number
of boxes that is a multiple of $N$.

The quantum states of the model can be constructed by starting with the
ground state $|0>$, annihilated by all operators $A_{jk}$
\be
A_{jk} |0> = 0
\ee
and acting with any number of creation operators:
\be
|j_1 , k_1 ; j_2 , k_2 ; \dots j_n k_n > =
A_{j_1 k_1}^\dagger A_{j_2 k_2}^\dagger \dots A_{j_n k_n}^\dagger |0>
\ee
For any $c-$number (classical) matrix $\Phi$, the commutation relations
\beqs
[ \tr (\Phi L) , A^\dagger ] &=& \Phi A^\dagger \cr
[ \tr (\Phi R) , \Ad ] &=& - A^\dagger \Phi \cr
[ \tr (\Phi J) , \Ad ] &=& \Phi \Ad - \Ad \Phi
\eeqs
imply that the first (left) index of the matrix 
$(A^\dagger )_{jk}$ transforms in the fundamental ($F$) under $L$, the second 
(right) index transforms in the antifundamental ($\bar F$) under 
$R$ and that $A^\dagger$ 
altogether transforms in the adjoint under $J$.
The general state transforms in a representation of the form
$F \times F \times \dots \times {\bar F} \times {\bar F} \times \dots$
with $n$ $F$s and $n$ $\bar F$s appearing, corresponding the the number
of free left and right indices in the state.

Physical states are chosen by imposing the constraint that $J$ is in
a fixed representation of $SU(N)$, say $r$. To do this, we start from
a generic state $|j_1 , k_1 ; j_2 , k_2 ; \dots j_n k_n >$ and contract
the free indices with Clebsh-Gordan coefficients that project it to this
representation. All states within the same representation $r$ are
gauge-equivalent and therefore represent a unique physical state.
States corresponding to inequivalent copies of $r$, however, contained
in the representation of the initial state correspod to
distinct physical states.

The above analysis can be done explicitly in the case of the representation
$r$ corresponding to the standard (spinless) Calogero model. Classically
the matrix commutator $i[M,{\dot M}] = J$ is of the form $\ell (v v^\dagger -1)$,
where $v$ is a vector of length squared equal to $N$. The representation $r_\ell$
corresponding to this is the fully symmetric one with a number of
Young tableau boxes equal to $\ell N$.

The easiest way to realize this is by quantizing the classical expression
$J = \ell v v^\dagger -\ell = \psi \psi^\dagger -\ell$ promoting the 
vector components $\psi_j \ell^\half v_j$ to yet another set of harmonic
oscillator creation operators
\be
[ \psi_j , \psi_k^\dagger ] = \delta_{jk}
\ee
In this way the operator
\be
(J_\psi )_{jk} = - \psi_k^\dagger \psi_j
\ee
realizes the $U(N)$ algebra, again in a Jordan-Wigner construction over the
antifundamental. Its representation content includes all fully symmetric
products of the antifundamental, each of them represented in the subspace
of fixed excitation number $n = \psi_j^\dagger \psi_j$. The restriction
of $J$ to the symmetric representation $r_\ell$ can be expressed by the
condition
\be
J + J_\psi + \ell = 0
\ee
This means that the representations carried by $J$ and $J_\psi$ must be
conjugate to each other so that their sum (composition) contain the
identity, which fixes the representation of $J$ to be a totally symmetric
one. The $c$-number term $\ell$ serves the purpose of subtracting
the trace of $J_\psi$, which is required by the tracelessness of $J$.
It also fixes the $U(1)$ part of $J_\psi$, since tracing the above
relation gives
\be
\tr J_\psi + \ell N = -\sum_j \psi_j^\dagger \psi_j + \ell N = 0
\ee
Wde recover the condition that $n=\ell N$ is the total number of
boxes in the Young tableau of $J_\psi$ and thus also of $J$, fully
fixing the desired representation.

As we stated earlier, representations of $J$ must have a number
of Young tableau boxes that is an integer multiple of $N$.
An important corollary of the above analysis, therefore, 
is that $\ell$ must be quantized to an integer. This is a new
feature of the Calogero system as deriving from the matrix
model. It is {\it not} a general requirement for the quantum
Calogero model, which is perfectly well-defined for fractional
values of $\ell$. The source of this quantization is the
enlargement of the symmetry group of the system from $S_N$ to
$SU(N)$ for the matrix model: the enlarged symmetry has a
global `anomaly' which requires the quantization of the coupling
constant $\ell$, in analogy with similar effects in gauge theory.

Finally, states of the theory can now be constructed in terms
of the vaccum state annihilated by $A$ and $\psi$ upon the
action of $\psi^\dagger$ and $A^\dagger$. The full state must 
be a singlet under $J + J_\psi$, which means that all indices, 
{\it including those of $\psi^\dagger$}, must be contracted.
Gauge invariant contraction of indices can be done by matrix
multiplication, but tracing or by multiplication of the vector
$\psi^\dagger$ and the matrix $A^\dagger$.

Each $\psi^\dagger$
in the state, therefore, will leave one index hanging, since there
is no way to contract it with another $\psi^\dagger$ and multiplication
with $A^\dagger$ still leaves one uncontracted index. The only way
to contract these indices is using the only invariant tensor of
$SU(N)$ that includes all fundamental indices, namely the $N$-fold
antisymmetric tensor $\epsilon_{j_1 \dots j_N}$. This means that
$\psi^\dagger$s must come in multiples of $N$, recovering once
more the condition that the total excitation number $n=\ell N$
must be a multiple of $N$. Further, operators contracted with 
$\epsilon$ must all be distinct, otherwise the product would
vanish due to antisymmetry.

The generic form of the physical states is \cite{HeRa}
\be
\left[ \tr \Ad \right]^{m_1} \left[ \tr (\Ad )^2 \right]^{m_2} \dots 
\left[ \tr (\Ad )^N \right]^{m_N} \left[
\epsilon_{j_1 \dots j_N} \psi_{j_1}^\dagger (\psi^\dagger \Ad)_{j_2}
\dots (\psi^\dagger (\Ad)^{N-1} )_{j_N} \right]^\ell |0>
\ee
Higher traces of $\Ad$ can be related to the first $N$ ones,
and other, more general ways of contracting operators with $\epsilon$
tensors can be reduced to linear combinations of the above states.

This is an eigenstate of the hamiltonian $H = \omega \tr (A^\dagger A)
+ \omega N^2 /2$ with its energy given by the total number of $\Ad$
oscillators appearing in the state plus a constant.
The $\psi$-dependent prefactor contributes an energy equal to
$\omega \ell N(N-1)/2$. We conclude that the energy spectrum is
\be
E = \omega \left( \sum_{k=1}^N m_k + \ell \frac{N(N-1)}{2} +
\frac{N^2}{2} \right)
\ee
We recognize the excitation energies as those of $N$ noninteraction
bosons in a harmonic oscillator potential,
expressed in terms of their collective excitations ($m_k$ represents
the energy gap between the top $k$ bosons and the next lower one
on the single-particle oscillator spectrum). By standard bosonization
arguments, the same excitation spectrum applies to a set of $N$
noninteracting fermions in a harmonic oscillator ($m_k +1)$ represents
the energy gap between the top $k$ fermions and the next lower one
on the single-particle oscillator spectrum). So we may rewrite the above
spectrum as
\be
E = \omega \left( \sum_{i=1}^N n_i + \ell \frac{N(N-1)}{2} +
\frac{N^2}{2} \right) =
\omega \left( \sum_{i_1}^N {\bar n}_i +
\frac{N}{2} \right)
\ee
where $n_1 \le n_2 \le \dots n_N$ are single-particle bosonic excitation
numbers while
the `pseudoexcitation' numbers ${\bar n}_i$ have been defined as
\be
{\bar n}_i = n_i + (\ell+1) (i-1)
\ee
We recover the spectrum of the Calogero model as exposed in a previous
section, with the extra shift $\ell \to \ell+1$. This is a quantum
shift of the `bare' parameter $\ell$ appearing in the classical matrix
model to the renormalized value $\ell +1$ entering the Calogero model.
In particular, the single sector $\ell = 0$ of the matrix model
corresponds to $\ell=1$ in the Calogero model, that is, fermions.
The `fermionization' of the eigenvalues of the matrix model due to the
quantuym mechanical measure arising out of integratig out the angular
variables of the matrix is a well-known effect. 

Other, more general representations of $J$ can be dealt with
in similar ways. As shall be explaind later, they correspond to Calogero
particles with internal degrees of freedom (`spin'). We shall examine
these cases in the context of the unitary matrix model. We have analyzed
quite enough already the hermitian matrix model and we should leave
something interesting for the unitary case!

\section{The unitary matrix model}
\subsection{Classical analysis}

The hermitian matrix model works well for particles on the line
but has trouble representing particles on periodic spaces. The most
natural candidate for such models would be a unitary $N \times N$ matrix
$U$. Its eigenvalues are phases and naturally live on the circle. We start,
therefore, with a lagrangian that represents the invariant kinetic energy
on the space of such matrices:
\be
\LL = -\half \tr ( U^{-1} {\dot U} )^2
\ee
A potential could in principle be included but we are interested in the
translationally invariant case and will omit it. The treatment is similar
as before, and we just summarize the relevant facts. 

The lagrangian is,
in fact, invariant under separate left- and right-multiplications of $U$ 
by time-independent unitary matrices and and there are two
corresponding conserved matrix angular momenta $L$ and $R$:
\begin{eqnarray}
U \to  VU : ~~~~~~~L &=& i {\dot U} U^{-1} \\
U \to  UW^{-1} :~~~  R &=& -i U^{-1} {\dot U}
\end{eqnarray}
The unitary conjugation that preserves the eigenvalues corresponds to
$W=V$ and its generator is
\be
J = L + R = i [ {\dot U} , U^{-1} ]
\ee
The rest of the discussion is as previously. Parametrizing 
\be
U = V \Lambda V^{-1} ~~{\rm with}~~
\Lambda = diag \{ e^{i x_i} , \dots e^{i x_N} \}
\label{UVparam}
\ee
the hamiltonian becomes, after a few steps,
\be
H = \sum_i \half p_i^2 + \half \sum_{i \neq j} \frac{
K_{ij} K_{ji}}{4 \sin^2 \frac{x_i - x_j}{2} }
\ee
where, as before,
\be
K = V^{-1}  J V
\ee
Choosing $J=K=0$ reproduces free particles on the circle, while
choosing $J = \ell (u u^\dagger -1)$ we obtain $K_{ij} K_{ji} = \ell^2$
and we recover the Sutherland inverse-sine-square model
\be
H = \sum_i \half {\dot x}_i^2 + \half \sum_{i \neq j} \frac{
\ell^2}{4 \sin^2 \frac{x_i - x_j}{2} }
\ee

This model is integrable and solvable by the same techniques as
the hermitian one. The conserved invariant quantities are
\be
I_n = \tr L^n = \tr (-R)^n = \tr (i U^{-1} {\dot U})^n
\ee
and the solution is
\be
U = B e^{iCt}
\ee
with $B$ a unitary and $C$ a hermitian matrix satisfying
\be
BCB^{-1} - C = J
\ee
For the Sutherland case with $J=\ell (u u^\dagger -1)$, $u_i =1$,
$B,C$ become
\be
B_{jk} = \delta_{jk} e^{i q_j} ~,~~~
C_{jk} = \delta_{jk} \, p_j + (1-\delta_{jk} ) \frac{i\ell}
{e^{i(q_j - q_k )} -1}
\ee
where, clearly, $q_i$ and $p_i$ are initial positions and momenta.

We conclude by mentioning that, upon scaling $x \to \alpha x$,
$t \to \alpha^2 t$, the Sutherland model goes over to the free Calogero model. 
This is the `infinite volume' limit.

\subsection{Unitary matrix model: quantization}

The quantization of the unitary matrix model can be performed in a way
practically identical to the hermitian model. Indeed, it should be obvious
that the matrices $L$ and $R$ defined in the two models have idential
properties and the whole analysis can be done in a parallel way. The
new element of the unitary model is that the hamiltonian is proportional
to the quadratic Casimir $I_2 = \tr L^2$, rather than the linear one, and
appropriate bases of states have to be found in the degenerate spaces
of $I_1 = \tr L$ that diagonalize also $I_2$.

We shall, however, give an independent treatment of the unitary model.
The reasons are primarily pedagogical: we shall use an explicit index
notation, rather than the neat $T_{12}$ calculus of the previous sections,
just to familiarize ourselves with the alternative. Further, we shall
make specific use of the connection of the $N \times N$ unitary model with
the group $U(N)$, taking advantage of (and making contact with) known
representation facts for this group.

We begin, again, by defining a canonical
momentum matrix conjugate to the `coordinate' $U$
\be
\Pi = \frac{\partial \LL}{\partial {\dot U}} = - U^{-1} {\dot U} U^{-1}
\ee
The Poisson brackets are
\be
\{ U_{jk} , \Pi_{lm} \} = \delta_{jm} \delta_{lk}
\label{UPi}
\ee
$\Pi$ is somewhat unpleasant, being neither unitary nor hermitian.
We prefer to work in terms of the hermitian matrices $L$ and $R$ defined
previously
\be
L = i {\dot U} U^{-1} = -iU\Pi ~,~~~
R = -i U^{-1} {\dot U} = i \Pi U
\ee
Using (\ref{UPi}) we derive the following Poisson brackets:
\begin{eqnarray}
\{ L_{jk} , L_{lm} \} &=& i ( L_{jm} \delta_{lk} - \delta_{jm} L_{lk} ) \\
\{ L_{jk} , R_{lm} \} &=& 0 \\
\{ R_{jk} , R_{lm} \} &=& i ( R_{jm} \delta_{lk} - \delta_{jm} R_{lk} ) 
\end{eqnarray}
The above is nothing but two copies of the $U(N)$ algebra in disguise.
To see this, expand the matrices $L$ and $R$ in the basis of the 
fundamental generators of $SU(N)$ $T^a$ plus the unit matrix:
\be
L = L^o + 2 \sum_{a=1}^{N^2 -1} L^a T^a 
\label{Loa}
\ee
\be
R = R^o + 2 \sum_{a=1}^{N^2 -1} R^a T^a
\label{Roa}
\ee
with $L^o$, $L^a$, $R^o$, $R^a$ numbers. Then use the $SU(N)$
commutation relations
\be
[ T^a , T^b ] = i f^{abc} T^c
\ee
as well as the normalization 
\be
\tr ( T^a T^b ) = \half \delta_{ab}
\ee
to show that the expansion coefficients satisfy the Poisson algebra
\begin{eqnarray}
\{ L^a , L^b \} &=&  f^{abc} L^c  \\
\{ L^a , R^b \} &=& 0 \\
\{ R^a , R^b \} &=&  f^{abc} R^c
\end{eqnarray}
while $L^o$, $R^o$ are central. Note that the $U(1)$ charges
\be
L^o = - R^o = \frac{1}{N} \tr (-i U^{-1} {\dot U}) = \frac{1}{N}
\sum_i {\dot x}_i
\ee
are essentially the total momentum of the system.

We are now ready to perform quantization. In the
$U$-representation, where states are functions of $U$, $\Pi$ becomes 
the matrix derivative $\Pi_{jk} = -i \delta_U$, acting as
\be
\delta_U \tr (UB) = B ~,~~~ \delta_U \tr (U^{-1} B) = - U^{-1} B U^{-1}
\ee
where $B$ is a constant matrix, and similarly on expressions containing
more $U$'s.
$L$ and $R$, upon proper ordering, are represented as
\be
L = -U \delta_U ~,~~~ R = \delta_U \, \cdot \, U
\ee
where in $R$ it is understood that we {\it first} act with the derivative
and {\it then} right-multiply the result by $U$. With this ordering,
$L$ and $R$ become the proper $U(N)$ operators acting as
\be
L \tr (UB) = -UB ~,~~~ L \tr (U^{-1} B) = B U^{-1} 
\ee
\be
R \tr (UB) = BU ~,~~~ R \tr (U^{-1} B) = - U^{-1} B 
\ee
It is also useful to express their action on arbitrary functions of $U$ as
\begin{eqnarray}
\tr (i\epsilon L) f(U) &= f((1-i\epsilon)U) - f(U) \\
\tr (i\epsilon R) f(U) &= f(U(1+i\epsilon)) - f(U) 
\end{eqnarray}
where $\epsilon$ is an arbitrary infinitesimal hermitian matrix,
emphasizing their role as generators of left- and right-multiplication
on $U$. Correspondingly, the operators $L^a$ and
$R^a$ satisfy the $SU(N)$ algebra. Their action can be obtained by taking
$\epsilon = \varepsilon T^a$ with $\varepsilon$ an infinitesimal 
scalar parameter, that is,
\begin{eqnarray}
i\varepsilon L^a f(U) &= f((1-i\varepsilon T^a) U) - f(U)
\label{elra} \\
i\varepsilon R^a f(U) &= f(U(1+i\varepsilon T^a)) - f(U) 
\end{eqnarray}
The hamiltonian, being classically
the kinetic term on the manifold of unitary matrices $U(N)$, quantum 
mechanically becomes the laplacian operator on the manifold \cite{NAM}.
Using (\ref{Loa},\ref{Roa}) it is expressed as
\be
H = \half \tr L^2 = \sum_a (L^a)^2 + \half N (L^o)^2 =
\sum_a (R^a)^2 + \half (R^o)^2 = \half \tr R^2 
\ee
It is, therefore, the common quadratic Casimir of the left- and
right-$SU(N)$ algebra plus the square of the $U(1)$ charge, the two
parts identifiable as the relative and center-of-mass energy respectively.

Quantum mechanical states grouping into irreducible representations
of the $L$ and $R$ $SU(N)$ algebras will, thus, be degenerate
multiplets of the hamiltonian. The $U(1)$ (center of mass) part 
trivially separates: we can boost any state by any desired total
momentum $NP$ by multiplying the wavefunction by $(\det U)^P$. We will
examine only the $SU(N)$ part from now on.

A natural basis of states for the Hilbert space are the matrix elements
of the unitary irreducible representations (irreps) of $SU(N)$. Let $R$ 
denote such an irrep, $R(U)$ the matrix that represents $U$ in this irrep
and $R_{\alpha \beta} (U)$ the $\alpha \beta$ matrix element of this matrix.
Clearly $\alpha$ and $\beta$ range from 1 to the dimensionality of
$R$, $d_R$. $R_{\alpha \beta} (U)$ are a complete orthonormal basis of
wavefunctions for $U$, that is
\be
\int [dU] R_{\alpha \beta} (U) R'_{\gamma \delta} (U)^* = \delta_{R R'}
\delta_{\alpha \gamma} \delta_{\beta \delta}
\ee
with $[dU]$ the volume element on the space of $SU(N)$ matrices as
implied by the metric $ds^2 = -\tr ( U^{-1} dU )^2$, also called the
Haar measure. 

We will, now, show that each $R_{\alpha \beta} (U)$ is an eigenstate
of the hamiltonian with eigenvalue equal to the quadratic Casimir of
$R$, $C_R$. Qualitatively, after the discussion of the last paragraphs,
this should be obvious: $L$ and $R$ generate the transformations
$U \to V^{-1} U$ and $U \to UW$. $R (U)$ transforms in the conjugate
irrep ${\bar R}$ under $L$ and in the irrep $R$ under $R$. Since $H$ 
is the common quadratic Casimir of $L$ and $R$ we conclude that all
$d_R^2$ states $R_{\alpha \beta} (U)$ are energy eigenstates with 
eigenvalue $C_R = C_{\bar R}$.

(If you are confused about $L$ generating 
$U \to V^{-1} U$ rather than $U \to VU$, think of
the difference between active and passive transformations, which is
relevant when shifting from classical to quantum: $\psi (x-a)$
shifts the wavefunction by $+a$. Also, although classical transformations
on $U$ compose properly,
\be
V_1 ( V_2 U) = (V_1 V_2 ) U
\ee
quantum mechanically the operators $\hat V$ that perform the shift
$U \to VU$ on the argument of the wavefunction would compose
\be
{\hat V}_1 ( {\hat V}_2 f(U)) = {\hat V}_1 f( V_2 U) 
= f( V_2 V_1 U) = ({\hat V}_2 {\hat V}_1) f(U)
\ee
Therefore we need to invert the action of $\hat V$ to get the right
composition law.

Let us prove the fact $H R_{\alpha \beta} (U) = C_R R_{\alpha \beta} (U)$
more analytically. Since $R (U)$ is a representation, it obeys the
group property
\be
R_{\alpha \beta} (UV) = \sum_\gamma 
R_{\alpha \gamma} R_{\gamma \beta} (V)
\ee
{}From (\ref{elra}) we have
\begin{eqnarray}
(1+i\varepsilon ) L^a R_{\alpha \beta} (U) &=& R_{\alpha \beta} 
((1-i\varepsilon T^a) U) = R_{\alpha \gamma} (1-i\varepsilon T^a)
R_{\gamma \beta} (U) \\
&=& R_{\alpha \beta} (U) - i\varepsilon
R_{\alpha \gamma}^a R_{\gamma \beta} (U)
\end{eqnarray}
where $R^a = R(T^a )$ is the $a$-th generator of $SU(N)$ in the
$R$ representation. So
\be
L^a R_{\alpha \beta} (U) = -R_{\alpha \gamma}^a R_{\gamma \beta} (U)
\ee
and
\be
\sum_a (L^a )^2 R_{\alpha \beta} (U) = \sum_a R_{\alpha \gamma}^a 
R_{\gamma \delta}^a R_{\delta \beta} (U) = 
\sum_a (R^a )_{\alpha \delta}^2 R_{\delta \beta} (U) 
\ee
The sum $\sum_a (R^a )^2$ appearing above is the quadratic
Casimir in the irrep $R$ and is proportional to the identity matrix
$\delta_{\alpha \delta}$. So, finally,
\be
H R_{\alpha \beta} (U) = C_R R_{\alpha \beta} (U)
\ee
Incidentally, the spectrum spanned by $C_R$ for all $R$ is nothing but
the spectrum of $N$ free fermions on the circle with the ground
state energy and the center-of-mass energy subtracted, where the 
lengths $R_i$ of the rows of the Young tableau of $R$ correspond 
to the ``bosonized'' fermion momenta
\be
p_i = R_i -i+1
\ee
and where the center-of-mass energy has been subtracted. The condition
$R_i \ge R_{i+1}$ for the rows amounts to the fermionic condition
$p_i > p_{i+1}$. The spectrum of the full matrix model, then, 
is identical to the free fermion one but with
{\it different degeneracies}.

We have, therefore, identified all energy eigenstates of the matrix model.
It remains to implement the quantum analog of the choice of angular
momentum $J$, identify the corresponding reduced quantum model, and pick
the subspace of states of the full model that belongs to the reduced
model.

$J$ obeys itself the $SU(N)$ algebra (it is traceless, no $U(1)$ charge).
As in the hermitian model, a choice of value for $J$ amounts 
to a choice of irrep $r$ for this algebra.
States within the same irrep are related by unitary transformations
of $U$ and give the same dynamics; they correspond to a unique physical state.
Since $J=L+R$,
we see that states transforming under $(L,R)$ in the $({\bar R}, R)$
irreps will transform in the ${\bar R} \times R$ under $J$. So, only
irreps $r$ that are contained in the direct product of two mutually
conjugate irreps can be obtained for $J$. This amounts to irreps $r$
with a number of boxes in their Young tableau that is an integer multiple
of $N$, just as in the hermitian model case. 
(To get a feeling of this, consider the case $N=2$. Then
$J$ is an orbital-like realization of the angular momentum through
derivatives of $U$ and clearly cannot admit spinor representations.)

We must, therefore, project the $d_R^2$ states in $R_{\alpha \beta} (U)$
to the subspace of states transforming as $r$ under $L+R$.
Call $G({\bar R}, \alpha ; R, \beta | r , \gamma )$ the Clebsch-Gordan 
coefficient that projects these states to the $\gamma$ state of $r$.
Then the relevant states for this model become
\be
\Psi_R ( U ) = \sum_{\alpha , \beta} R_{\alpha \beta} (U) G({\bar R}, \alpha;
R, \beta | r , \gamma ) 
\ee
The index $\gamma$ labeling the states within $r$, as we argued before,
counts the $d_r$ gauge copies and does not imply a true degeneracy of states.
The degeneracy of the states produced by each $R$ is, then, given by the
number of times that the irrep $r$ is contained in the direct
product ${\bar R} \times R$ or, equivalently, the
number of times that $R$ is contained in $R \times r$. Calling this 
integer $D( R,r ; R )$, we obtain for the spectrum and degeneracies:
\be
E_R = C_R ~,~~~ D_R = D( R,r ; R )
\label{spectrum}
\ee
In particular, if $D_R =0$ the corresponding energy level is
absent from the spectrum.

Concluding, we mention that an approach which also reproduces the
spectrum and states of the Sutherland model is two-dimensional
Yang-Mills theory on the circle \cite{GN,MPYM}. This approach
is essentially equivalent to the matrix model above and we
will not be concerned with it.

\subsection{Reduction to spin-particle systems}

So we have derived the spectrum, degeneracy and wavefunctions of the
matrix model restricted to the sector $J=r$. Classically these
restrictions represented free particles ($J=0$), Sutherland particles
($J= \ell (v v^\dagger -1)$) or something more general. What are the
corresponding quantum systems?

To find these, let us reproduce here the expression of the reduced
hamiltonian in one of these sectors:
\be
H = \sum_i \half p_i^2 + \half \sum_{i \neq j} \frac{
K_{ij} K_{ji}}{4 \sin^2 \frac{x_i - x_j}{2} } -E_o
\label{Hquant}
\ee
This expression remains valid quantum mechanically upon a proper
definition (ordering) of the operator $K$. The only residual quantum
effect is a constant term $E_o$ that comes from the change of measure
from the matrix space to the space of eigenvalues.

Let us expand a bit on this without entering too deeply into the 
calculations. (For details se, e.g., \cite{Meh}.) The Haar measure 
in terms of the diagonal and angular part of $U$ has the form
\be
[dU] = \Delta^2 \, [dV]
\ee
where $[dV]$ is the Haar measure of $V$ and $\Delta$ is the 
Vandermonde determinant
\be
\Delta = \prod_{i<j} 2 \sin \frac{x_i - x_j }{2}
\ee
To see this, write the `line element' $-\tr (U^{-1} dU)^2$ in terms of
$V$ and $x_i$ using (\ref{UVparam}) and obtain
\be
-\tr (U^{-1} dU)^2 = \sum_i dx_i^2 - \sum_{i,j}
4 \sin^2 \frac{x_i - x_j }{2} (V^{-1} dV)_{ij} (V^{-1} dV)_{ji} 
\ee
This metric is diagonal in $dx_i$ and $(V^{-1} dV)_{ij}$. The 
square root of the determinant
of this metric, which gives the measure (volume element) on the space,
is clearly $\Delta^2$ times the part coming from $V$ which is the
standard Haar measure for $V$. (We get {\it two} powers of $4 \sin^2
\frac{x_i - x_j}{2}$ in the determinant, one from the real and one
from the imaginary part of $(V^{-1} dV)_{ij}$, so the square root
of the determinant has one power of $\Delta^2$.)

To bring the kinetic $x_i$-part into a `flat' form (plain second
derivatives in $x_i$) we must multiply the wavefunction with the
square root of the relevant measure (compare with the change from
cartesian to spherical coordinates in central potential problems).
The net result is that the wavefunction $\Psi$ in terms of $x_i$ and
$V$ is the original wavefunction $\psi (U)$ of the matrix model
times the Vandermonde determinant. This, however, also produces
an additive constant $E_o$ which comes from the action of the entire
$x_i$-kinetic operator on $\Delta$. Noticing that $\Delta$ is 
nothing but the ground state wavefunction of $N$ free fermions on
the circle, we see that $E_o$ is the relevant fermionic ground 
state energy
\be
E_o = \frac{N(N^2 -1)}{24}
\ee
This is the famous `fermionization' of the eigenvalues produced by
the matrix model measure.

To determine the proper ordering for $K$ we examine its properties
as a generator of transformations. Since $U = V \Lambda V^{-1}$, 
and $J$ generates $U \to V' U V'^{-1} = (V' V) \Lambda (V' V)^{-1}$,
we see that $J$ generates left-multiplications of the angular part
$V$ of $U$. $K = V^{-1} J V$, on the other hand, generates {\it
right}-multiplications of $V$, as can be see from its form or
by explicit calculation through its Poisson brackets. As a result,
it also obeys the $SU(N)$ algebra. Its proper quantum definition, then,
is such that it satisfies, as an operator, the $SU(N)$ algebra.
It clearly commutes with the diagonal part $x_i$ and its momentum $p_i$,
since it has no action on it. Its dynamics are fully determined
by the hamiltonian (\ref{Hquant}) and its $SU(N)$ commutation relations.

We can, therefore, in the context of the particle model (\ref{Hquant}),
forget where $K$ came from and consider it as an independent set of 
dynamical $SU(N)$ operators. $K$, however, obeys some constraints.
The first is that, as is obvious from $K = V^{-1} J V$, $K$ carries 
{\it the same irrep $r$} as $J$. The second is subtler: a 
right-multiplication of $V$ with a diagonal matrix will clearly leave
$U = V \Lambda V^{-1}$ invariant. Therefore, this change of
$V$ has no counterpart on the `physical' degrees of freedom of the
model and is a gauge transformation. As a result, we get the `Gauss' law'
that physical states should remain invariant under such transformations.
Since $K$ generates right-multiplications of $V$, and $K_{ii}$ (no sum)
generates the diagonal ones, we finally obtain
\be
({\rm no~sum})~~K_{ii} = 0 ~~~({\rm on~physical~states})
\label{Kcon}
\ee
(A more pedestrian but less illuminating way to see it is: $J = i
[ U^{-1} , {\dot U} ]$, being a commutator, vanishes when sandwiched 
between the same eigenstate of $U$. Since $K$ is essentially $J$ in
the basis where $U$ is diagonal, its diagonal elements vanish.)
Note that the constraint (\ref{Kcon}) is preserved by the hamiltonian
(\ref{Hquant}).

The above fully fixes the reduced model Hilbert space as the product of
the $N$-particle Hilbert space times the $d_r$-dimensional space of $K$,
with the constraint (\ref{Kcon}) also imposed. The further casting of the
model into something with a more direct physical interpretation relies
upon a convenient realization of $K$. {\it Any} such realization will
do: simply break the representation of $SU(N)$ that it carries into
irreps $r$ and read off the spectrum for each $r$ from the results of
the previous section.

We shall implement $K$ in a construction {\it \`a la} Jordan-Wigner,
as exposed in the hermitian matrix model. Let 
$a_{mi}$, $a_{mi}^\dagger$, $m=1, \dots q$, $i=1, \dots N$ be a set of 
$Nq$ independent bosonic oscillators \cite{MPYM}:
\be
[ a_{mi} , a_{nj}^\dagger ] = \delta_{mn} \delta_{ij}
\ee
Then
\be
K^a = \sum_{m=1}^q a_{mi} T_{ij}^a a_{mj}
\ee
is a realization of the $SU(N)$ algebra. ($T_{ij}^a$ are the matrix
elements of the fundamental generators $T^a$.)
The corresponding matrix elements of $K$ are
\be
K_{ij} = \sum_{m=1}^q \left\{ a_{mi}^\dagger a_{mj} - \frac{1}{N} 
\left( \sum_k a_{mk}^\dagger a_{mk} \right) \delta_{ij} \right\}
\label{Kij}
\ee
Correspondingly, the coefficient of the Sutherland potential in 
(\ref{Hquant}) is (for $i \neq j$)
\be
K_{ij} K_{ji} = \sum_{m,n} a_{mi}^\dagger a_{ni} \, a_{nj}^\dagger 
a_{mj} + \sum_m a_{mi}^\dagger a_{mi}
\ee
We already see that the degrees of freedom of $K$ are redistributed into 
degrees of freedom for each particle in the above. Specifically, defining
\be
S_{i,mn} = a_{mi}^\dagger a_{ni} - \frac{1}{q} \left( \sum_{s=1}^q 
a_{si}^\dagger a_{si} \right) \delta_{mn}
\label{Smn}
\ee
and comparing with (\ref{Kij}) we see that the $S_i$ are $N$ independent
sets of operators each satisfying the $SU(q)$ algebra. Before expressing
$K_{ij} K_{ji}$ in terms of the $S_i$ let us see what the constraint
(\ref{Kcon}) implies:
\be
K_{ii} = \sum_{m=1}^q a_{mi}^\dagger a_{mi} - 
\frac{1}{N} \sum_{m,k} a_{mk}^\dagger a_{mk} = 0
\ee
$\sum_{m,k} a_{mk}^\dagger a_{mk}$ commutes with all $K_{ij}$
and all $S_{i,mn}$. It is, therefore, a Casimir and can be chosen as a 
fixed integer $\ell N$ equal to the total number operator of the subspace 
of the oscillator Fock space in which the model lives. The above
constraint, then, implies
\be
\sum_{m=1}^q a_{mi}^\dagger a_{mi} = \ell 
\label{Acon}
\ee
(We see why we had to choose the total number operator to be a
multiple of $N$: the operator in (\ref{Acon}) above is also a
number operator and can have only integer eigenvalues.)
Using this in (\ref{Smn}) we can express 
\be
a_{mi}^\dagger a_{ni} = S_{i,mn} + \frac{\ell}{q} \delta_{mn}
\ee
and therefore
\be
K_{ij} K_{ji} = \sum_{mn} S_{i,mn} S_{j,nm} + \frac{\ell (\ell+q)}{q}
= {\vec S}_i \cdot {\vec S}_j + \frac{\ell (\ell+q)}{q}
\ee
where ${\vec S}_i \cdot {\vec S}_i = \tr (S_i S_j )$ is the
$SU(q)$-invariant scalar product of the two $SU(q)$ `vectors.'
We finally obtain the hamiltonian as \cite{MPYM}
\be
H = \sum_i \half p_i^2 + \half \sum_{i \neq j} \frac{
2{\vec S}_i \cdot {\vec S}_j + \frac{\ell (\ell +q)}{q}}
{4 \sin^2 \frac{x_i - x_j}{2} }
\label{HSquant}
\ee
So it is a Sutherland-like model, in which the particles also
carry $SU(q)$ internal degrees of freedom (`spins') and the potential
contains a pairwise antiferromagnetic interaction between the spins.

It remains to specify the representation in which the $SU(q)$ spins
are and find the irreps contained in this realization of $K$, therefore
obtaining the spectrum. A realization of the form (\ref{Kij}) for
$q=1$ in terms of bosonic oscillators contains all {\it totally symmetric}
irreps of $S(N)$ (that is, the ones with a single row in their Young
tableau). (\ref{Kij}) is essentially the direct product of $q$ such
independent realizations, so it contains all direct products of $q$
totally symmetric irreps. This produces all irreps with up to $q$
rows in their Young tableau, perhaps more than once each. The constraint
(\ref{Acon}), however, implies that the total number of boxes in the
Young tableau of these irreps is $\ell N$. We recover once more the
constraint that we derived before based on the origin of $r$ as a
component of ${\bar R} \times R$.

Similarly, the realization (\ref{Smn}) of $S_i$ contains all the
totally symmetric irreps of $SU(q)$. (\ref{Acon}) implies that the number
of boxes of these irreps is equal to $\ell$, so the spins $S_i$ are each
in the $\ell$-fold symmetric irrep of $SU(q)$. Solving this model
amounts to decomposing the tensor product of these $N$ spins into 
irreducible components of $SU(q)$. Each such component corresponds to
a subspace of the Hilbert space with a fixed total spin $S$. This same 
irrep, interpreted as an irrep $r$ of $SU(N)$, will be the corresponding
irrep of $K$, and also of $J$, and thus will determine the spectrum of 
this sector through (\ref{spectrum}).

Let us elucidate the above by reproducing the two simplest cases: free 
particles and (spinless) Sutherland particles, comparing with the classical
treatment. 

a) Free particles correspond to $J=K=0$. So there is no spin and no potential
and we have non-interacting particles. From (\ref{spectrum}) we see that all
$D_R$ are one, and thus the spectrum is the free fermion one, as commented
before. The matrix model naturally quantizes free particles as fermions.

b) Spinless Sutherland particles correspond, classically, to $J = \ell
(v v^\dagger -1)$. So $J$ is rank one (ignoring the trace). Quantum
mechanically this corresponds to the irrep $r$ of $J$ having only one
row and therefore only one independent Casimir. Since $q$ in the realization
above corresponds to the number of rows, we must have $q=1$. Spins,
therefore, are absent.
The strength of the potential becomes $\ell (\ell +1)$ where $\ell N$ is
the number of boxes in the one row of $r$. By standard Young tableau
rules we see that the degeneracy $D_R$ is one if the row lengths of $R$
satisfy
\be
R_i \ge R_{i+1} + \ell
\ee
else it is zero. The spectrum of this model is, then, the same as the
spectrum of free particles but with the selection rule for their momenta
\be
p_i \ge p_{i+1} + \ell +1
\ee
We recover the `minimum distance' selection rule of the Calogero model that
led to the interpretation as particles with generalized statistics! Only,
in this case, the statistics parameter $\ell +1$ is a positive integer.
The wavefunctions of the Sutherland model can be related to characters of
the $U(N)$ group (see, e.g., \cite{FGP}).

A Jordan-Wigner realization of $K$ in terms of
{\it fermionic} oscillators is also useful and leads to particles with spins
interacting via {\it ferromagnetic} Sutherland-type potentials. The
hamiltonian becomes \cite{MPYM}
\be
H = \sum_i \half p_i^2 - \half \sum_{i \neq j} \frac{
2{\vec S}_i \cdot {\vec S}_j + \frac{\ell(\ell -q)}{q}}
{4 \sin^2 \frac{x_i - x_j}{2} }
\label{HFquant}
\ee
where now the spins are in the $\ell$-fold {\it anti}symmetric irrep of
$SU(q)$. We will not elaborate further and leave the details as an exercise
to the reader.

Clearly there are other, more general realizations, involving a mixture
of fermionic and bosonic oscillators. Such a construction, e.g., involving a
single bosonic and a single fermionic oscillator produces the so-called
supersymmetric Calogero model \cite{AAAP}. The model of \cite{Park} also
falls into this category.

We close this section by noting that the unitary matrix model can be
thought as motion on a space of constant positive curvature $U(N)/G$,
where $G$ is the symmetry group that leaves the angular momentum $J$
(with which we reduce the system) invariant. For the spinless Sutherland
model, we have $G=U(N-1) \times U(1)$, where $U(N-1)$ rotates in the
$N-1$ directions normal to the vector $v$ and $U(1)$ is the phase
arbitrariness of $v$. The spin systems correspond to more general
choices of $G$. A similar construction for spaces of constant negative
curvature lead to systems with a hyperbolic sine-squared
interaction \cite{neg}. Such hyperbolic systems can be easily obtained
by analytically continuing the variables $x_i$ and $p_i$ to the
imaginary axis, which leaves the hamiltonian real.

\subsection{Concluding remarks about the matrix model approach}

In conclusion, the matrix model has provided us with the following:

1. An augmentation of the permutation group into the $SU(N)$ group
and a corresponding possibility to define statistics through the irreps
of $SU(N)$.

2. A realization of generalized scalar statistics but with {\it quantized}
statistics parameter $\ell +1$ in terms of the Calogero model.

3. A realization of generalized `non-abelian statistics' in terms of
particles with internal degrees of freedom interacting through a generalized
Calogero-type potential.

4. A systematic way of solving the above models.

\noindent
What the matrix models has {\it not} provided is

1. A realization of the Calogero model for fractional values of the
interaction parameter $\ell$.

2. A realization of spin-Calogero systems with the spins in arbitrary 
(non-symmetric) representations.

3. A control of the coupling strength of the potential for the spin-Calogero
models. (Note that the coefficient of ${\vec S_i} \cdot {\vec S_j}$ terms
is fixed to $\pm 2$ and also the constant therm $\ell (\ell +q)$ is
entirely fixed by the spin representation.)

There exist generalizations of the above models both at the classical
\cite{BL,PolR} and the quantum level \cite{AP}. They all share,
however, the limitations spelled out above, especially (3).
These restrictions are important, in the quest of more general
statistics but also from the more practical point of view of solving
spin-chain models with spins not in the fundamental representation, as we
will shortly explain. For this reason, yet a different approach will be
pursued in the next section, namely, the operator approach.

\section{Operator approaches}

The matrix model connection provided us with a powerful tool
that not only allowed us to generalize the notion of identical
particle but also
led to the full quantum solution of a set of spin-generalized
Calogero models.

As noted, however, in the conclusion of the preceding lecture,
the matrix model fixes the coefficient of the spin-interaction
and scalar interaction terms to $\pm 2$ and $\pm \ell (\ell \pm q)$
respectively. We cannot choose these coefficients at will.

We would like to have an approach that defeats this restriction
and leads to spin models with arbitrary coupling strengths. (This
is necessary to attack spin-chain systems through the 
infinite-coupling limit trick to be explained later.) Such an approach
should also be able to bypass the excursion to matrix models and 
deal more directly with these systems in an algebraic way.
This will be achieved with the exchange operator formalism \cite{PolEX}.

\subsection{Exchange operator formalism}

Consider the operators $M_{ij}$ that permute the {\it coordinate} 
degrees of freedom of $N$ particles in one dimension which could,
in principle, also have internal degrees of freedom ($M$ for {\it
metathesis}, to avoid confusion with momenta $p_i$). They satisfy
the permutation algebra (symmetric group), in particular
\begin{eqnarray}
&M_{ij} = M_{ij}^{-1} = M_{ij}^\dagger = M_{ji} \\
&[ M_{ij} , M_{kl} ] = 0 ~~~{\rm if}~i,j,k,l~{\rm distinct} \\
&M_{ij} M_{jk} = M_{ik} M_{ij} ~~~{\rm if}~i,j,k~{\rm distinct}
\end{eqnarray}
Any operator $A_i$ on the phase space satisfying
\begin{eqnarray}
M_{ij} A_k &=& A_k M_{ij} ~~~{\rm if}~i,j,k~{\rm distinct}\\
M_{ij} A_i &=& A_j M_{ij}
\end{eqnarray}
will be called a {\it one-particle} operator (even though
it may involve the coordinates and momenta of many particles).

We construct the following one-particle operators \cite{PolEX}:
\be
\pi_i = p_i + \sum_{j \neq i} i W(x_i - x_j ) M_{ij} \equiv
p_i + \sum_{j \neq i} i W_{ij} M_{ij} 
\ee
We shall view the $\pi_i$ as generalized momenta. To ensure their
hermiticity the {\it prepotential} $W(x)$ should satisfy
\be
W(-x) = -W(x)^*
\ee
We shall construct the corresponding `free' hamiltonian from $\pi_i$
\be
H = \sum_i \half \pi_i^2
\ee
In terms of the original $p_i$ this hamiltonian will, in general
contain linear terms. To ensure that such terms are absent we must
further impose
\be
W(-x) = -W(x) = {\rm real}
\ee
With the above restriction the hamiltonian $H$ and commutation relations
of the $\pi_i$ become
\be
[ \pi_i , \pi_j ] = \sum_k W_{ijk} (M_{ijk} - M_{jik} )
\ee
\be
H = \sum_i \half p_i^2 + \sum_{i<j} \left( W_{ij}^2 + W_{ij}' M_{ij}
\right) + \sum_{i<j<k} W_{ijk} M_{ijk}
\ee
where we defined the three-body potential and cyclic permutation
\begin{eqnarray}
W_{ijk} &=& W_{ij} W_{jk} + W_{jk} W_{ki} + W_{ki} W_{ij} \\
M_{ijk} &=& M_{ij} M_{jk} 
\end{eqnarray}
To obtain an interesting and tractable model, $W_{ijk}$, which
appears in the commutator $[\pi_i , \pi_j ]$ and also as a three-body
potential, should vanish or at most be a constant. This leads to
a functional equation for $W(x)$:
\be
W(x) W(y) - W(x+y) \left[ W(x)+W(y) \right] = {\rm const} (= W_{ijk} )
\ee
We present the solutions:

a) $W_{ijk} = 0 ~\rightarrow~ W(x) = \ell /x$

b) $W_{ijk} =-\ell^2 <0 ~\rightarrow~ W(x) = \ell \cot x$

c) $W_{ijk} = +\ell^2 >0 ~\rightarrow~ W(x) = \ell \coth x$

\noindent
Let's examine each case.

a) In this case the $\pi_i$ become
\be
\pi_i = p_i + \sum_{j \neq i} \frac{i\ell}{x_{ij}} M_{ij} 
\ee
and satisfy
\be
[ \pi_i , \pi_j ] =0
\ee
The $\pi_i$ commute, so we can consider them as independent momenta.
(They are sometimes referred to as Dunkl operators \cite{Dunk}.)
The hamiltonian reads
\be
H = \sum_i \half p_i^2 + \sum_{i<j} \frac{\ell (\ell -M_{ij} )}
{x_{ij}^2}
\ee
We obtain a Calogero-like model with {\it exchange interactions}.
Yet it is nothing but a free model in the commuting momenta $\pi_i$.
Integrability is immediate: the permutation-invariant quantities
\be
I_n = \sum_i \pi_i^n
\ee
obviously commute with each other. If we assume that the particles
carry no internal degrees of freedom and are bosons or fermions then
$M_{ij} = \pm 1$ on physical states. The model becomes the standard
Calogero model and we have proved its integrability in one scoop.
(You may be left with a question mark: the hamiltonian and the
other integrals $I_n$ become the standard Calogero ones if $M_{ij}
= \pm 1$, so these reduced integrals will commute on the bosonic or 
fermionic subspace; but will they also commute on the {\it full}
Hilbert space? Prove for yourself that this is indeed the case.)

We can also construct harmonic oscillator operators \cite{PolEX,BHV}.
The commutators between $x_i$ and $\pi_i$ are
\begin{eqnarray}
{[}x_i , \pi_i {]} &=& i \left( 1+ \ell \sum_{j \neq i}  M_{ij} 
\right) \\
{[}x_i , \pi_j {]} &=& -i\ell M_{ij} ~~~ (i \neq j)
\end{eqnarray}
Defining
\begin{eqnarray}
a_i &=& \frac{1}{\sqrt 2} \left( \pi_i - i\omega x_i \right) \\
a_i^\dagger &=& \frac{1}{\sqrt 2} \left( \pi_i + i\omega x_i \right) 
\end{eqnarray}
we can show
\begin{eqnarray}
{[} a_i , a_i^\dagger {]} &=& \omega \Bigl( 1+\ell \sum_{j \neq i} 
M_{ij} \Bigr) \\
{[} a_i , a_j^\dagger {]} &=& -\omega \ell M_{ij} ~~~(i \neq j) \\
{[} a_i , a_j {]} &=& {[} a_i^\dagger , a_j^\dagger {]} = 0
\end{eqnarray}
This is an extended version of the Heisenberg algebra involving the
permutation operators. The corresponding oscillator hamiltonian reads
\be
H = \sum_i \half ( a_i^\dagger a_i + a_i a_i^\dagger ) =
\sum_i \half p_i^2 + \sum_i \half \omega^2 x_i + \sum_{i<j}
\frac{\ell (\ell -M_{ij} )}{x_{ij}^2}
\ee
and satisfies
\be
[ H , a_i ] = \omega a_i ~,~~~ [ H , a_i^\dagger ] =\omega a_i^\dagger
\label{Hacomm}
\ee
This is the harmonic Calogero model with exchange interactions,
which becomes again the standard model on bosonic or fermionic subspaces
for particles without internal degrees of freedom. Since
\be
H = \sum_i a_i^\dagger a_i + \half N \omega + \half \ell \omega 
\sum_{i \neq j} M_{ij}
\ee
we see that on bosonic or fermionic spaces the state annihilated by all
$a_i$ (if it exists) will be the ground state. Solving $a_i \psi =0$
we obtain for the ground state wavefunction
\begin{eqnarray}
\psi_B &=& \prod_{i<j} |x_{ij} |^\ell e^{-\half \omega \sum_i x_i^2} \\
\psi_F &=& \prod_{i<j} \left\{ sgn(x_{ij} ) |x_{ij} |^{-\ell} 
\right\} e^{-\half \omega \sum_i x_i^2} 
\label{BFgr}
\end{eqnarray}
For $\ell >0$ the bosonic state is acceptable, while for $\ell <0$
the fermionic one is acceptable. In the ``wrong'' combinations of
statistics and sign of $\ell$ the ground state is not annihilated
by the $a_i$, but it is still annihilated by all permutation-invariant
combinations of the $a_i$. 

From (\ref{Hacomm}) we see that we can find the spectrum of this model
for fermions or bosons by acting on the ground state with all
possible permutation-symmetric homogeneous polynomials in the 
$a_i^\dagger$. A basis bor these is, e.g.,
\be
A_n = \sum_i (a_i^\dagger )^n
\ee
So the spectrum is identical to non-interacting fermions or bosons,
but with a different ground state energy. For the `right' combinations
of $\ell$ and statistics, where (\ref{BFgr}) are the correct ground
state wavefunctions, the ground state energy is
\be
E_o = \frac{N}{2} \omega + \frac{N(N-1)}{2} |\ell | \omega
\ee
which is the correct Calogero result.

Finally, the quantities 
\be
I_n = \sum_i h_i^n = \sum_i (a_i^\dagger a_i )^n
\ee
can be shown to commute \cite{PolEX}, and therefore this system is 
also integrable.
It is left as an exercise to find the commutation relations of the $h_i$
and show that $[I_n ,I_m ] =0$.

b) In the case $W(x) = \ell \cot x$ we have 
\be
\pi_i = p_i + i \cot x_{ij} M_{ij}
\ee
\be
[ \pi_i , \pi_j ] = - \ell^2 \sum_k ( M_{ijk} - M_{jik} )
\ee
so the momenta are now coupled. The hamiltonian becomes
\be
H = \sum_i \half p_i^2 + \sum_{i<j} \frac{\ell (\ell -M_{ij} )}
{\sin^2 x_{ij}} - \ell^2 \left( \frac{N(N-1)}{2} + \sum_{i<j<k}
M_{ijk} \right)
\ee
We obtain the Sutherland model with exchange interactions plus
an extra term. On bosonic or fermionic states this becomes an
overall constant and we recover the standard Sutherland model.
Again, since $H$ is by construction positive definite, if a state
satisfying $\pi_i \psi =0$ exists it will be the ground state.
We obtain
\begin{eqnarray}
\psi_B &=& \prod_{i<j} |\sin x_{ij} |^\ell \\
\psi_F &=& \prod_{i<j} sgn(x_{ij}) |\sin x_{ij} |^\ell 
\end{eqnarray}
which are acceptable for the same combinations of $\ell$ and
statistics as before. For both cases $M_{ijk} =1$ so
\be
E_o = \ell^2 \frac{N(N^2 -1)}{24}
\ee
is the correct Sutherland model ground state energy. The excited
states can again be obtained in a (rather complicated) algebraic way
\cite{LV}. Finally, the quantities
\be
{\tilde \pi}_i = \pi_i + \ell \sum_{j \neq i} M_{ij} = p_i + 
e^{ix_i} \sum_{j \neq i} \frac{2\ell}{e^{ix_i} - e^{ix_j}} M_{ij}
\ee
can be shown to have the same commutation relations as the $h_i$
defined previously for the harmonic system. Therefore, the integrals
constructed from them
\be
I_n = \sum_i {\tilde \pi}_i^n
\ee
commute and the model is integrable.

c) For $W(x) = \ell \coth x$ we have a similar commutation
relation and a hamiltonian
\be
H = \sum_i \half p_i^2 + \sum_{i<j} \frac{\ell (\ell -M_{ij} )}
{\sinh^2 x_{ij}} + \ell^2 \left( \frac{N(N-1)}{2} + \sum_{i<j<k}
M_{ijk} \right)
\ee
This is the inverse-hyperbolic-sine-square model and supports
only scattering states. Its integrability can be obtained as
for the Sutherland model above, or simply as an `analytic 
continuation' of that model for imaginary period of space.
We will not examine it any further.

In conclusion, an exchange-family of models was introduced, solved and
related to the standard Calogero models in spaces of definite symmetry.
It is remarkable that all these proofs work directly, and only, at
the quantum domain (there is no classical analog of $M_{ij}$).

\subsection{Systems with internal degrees of freedom}

We can easily extend the previous results for particles with
internal degrees of freedom. For this, assume that the particles are
{\it distinguishable} or, equivalently, that they carry a number $q$
of (discrete) internal degrees of freedom (species) that can be used to 
(partially) distinguish them. Their states are spanned by $|x,\sigma>$,
where $\sigma=1, \dots q$ counts internal states. The total permutation
operator $T_{ij}$, then is 
\be
T_{ij} = M_{ij} \sigma_{ij}
\ee
where $\sigma_{ij}$ is the operator that permutes the internal states
of particles $i$ and $j$.

Let us, then, simply take states that are bosonic or fermionic
under total particle exchange: $T_{ij} = \pm 1$. On such states
\be
M_{ij} = \pm \sigma_{ij}
\ee
and the Calogero and Sutherland exchange model hamiltonians become
\cite{MPSC}
\be
H_c = \sum_i \half p_i^2 + \sum_i \half \omega^2 x_i + \sum_{i<j}
\frac{\ell (\ell \mp \sigma_{ij} )}{x_{ij}^2} 
\ee
\be
H_s = \sum_i \half p_i^2 + \sum_{i<j} \frac{\ell (\ell \mp\sigma_{ij} )}
{\sin^2 x_{ij}} - \ell^2 \left( \frac{N(N-1)}{2} + \sum_{i<j<k}
\sigma_{ijk} \right)
\ee
We get the Calogero and Sutherland models with spin-exchange
interactions. From the completeness
relation for the fundamental $SU(q)$ generators $T^a$
\be
\sum_{a=1}^{q^2 -1} T_{\alpha \beta}^a T_{\gamma \delta}^a = 
\half \delta_{\alpha \delta} \delta_{\gamma \beta}
-\frac{1}{2q} \delta_{\alpha \beta} \delta_{\gamma \delta}
\ee
we deduce the form of the operators $\sigma_{ij}$
\be
\sigma_{ij} = 2 {\vec S}_i \cdot {\vec S}_j + \frac{1}{q}
\ee
where $S_i^a$ acts as $T^a$ on the internal states of particle $i$.
So the spin-dependent interaction coefficient of the potential 
in the hamiltonian takes the form \cite{HH,Kaw,MPSC,HW}
\be
\mp \ell \left( 2{\vec S}_i \cdot {\vec S}_j \mp \ell +
\frac{1}{q} \right)
\ee
We have recovered the ferromagnetic and antiferromagnetic spin
model of the previous section but with {\it arbitrary} coefficient!
On the other hand, the spins are necessarily in the fundamental
of $SU(q)$. So we have obtained a generalization of the coupling
constant with respect to the matrix model but a restriction of the
allowed spins. 

Note that $\ell$ here is an arbitrary parameter, while $\ell$ in
(\ref{HSquant}) was the size of the symmetric representation of $S_i$.
For $\ell=1$ and spins in the fundamental, the matrix model and 
exchange-operator model agree. It is interesting to note that we 
can go from ferromagnetic to antiferromagnetic interactions either 
by changing the sign of $\ell$ or by changing the statistics
of the particles.

The solution of the above models can be obtained algebraically.
For the spin-Sutherland model this is rather complicated and is
related to the so-called Yangian symmetry \cite{Hetc,BGHP,Cher,AJ}. 
For the spin-Calogero model it is easier \cite{PolD}. 
Let us concentrate on the model with 
interaction $\ell (-2{\vec S}_i \cdot {\vec S}_j \mp \ell -
\frac{1}{q} )$ and define the operators
\be
A_n^\dagger = \sum_i (a_i^\dagger )^n ~,~~~
(A_n^a )^\dagger = \sum_i (a_i^\dagger )^n S_i^a
\ee
and their hermitian conjugates.
These form a complete set for all permutation-symmetric creation
and annihilation operators for all species of particles. Yet the
commutators 
among themselves and with $H$ do not involve $\ell$. They create,
therefore, the same spectrum of excitations over the ground state
as $N$ noninteracting bosons or fermions with $q$ species.
For $\ell >0$ the ground state is the bosonic one:
\be
\psi_B = \prod_{i<j} |x_{ij} |^\ell e^{-\half \omega \sum_i x_i^2}
\chi_s (\{ \sigma_i \})
\ee
where $\chi_s$ is a totally symmetric state in the $\sigma_i$.
The set of all $\chi_s$ forms the $N$-fold symmetric irrep
of the total spin $S = \sum_i S_i$. Therefore the ground state 
is $(N+q-1)! / N! (q-1)!$ times degenerate. For $\ell<0$ the
above is not normalizable any more. But we remember that we can
obtain the same model by starting from fermions and the opposite 
coupling $-\ell >0$. The ground state, then, is of a fermionic
type
\be
\psi_F = \sum_P (-1)^P 
\left(\prod_i \delta_{\sigma_i ,\alpha_i} \right)
\prod_{i<j} |x_{ij} |^{-\ell} x_{ij}^{\delta_{\alpha_i ,\alpha_j}} 
e^{\half \omega \sum_i x_i^2}
\ee
where $P$ are total particle permutations and $\alpha_i$ are a
set of fixed values for the indices $\sigma_i$ that determine the
state. Clearly the ground state will be obtained for the minimal
total power of $x_i$ appearing above, and that will happen for
a maximally different set of values $\alpha_i$. These states form
the $n$-fold antisymmetric irrep of the total spin $S$, where $n
= N(mod q)$. The ground state is, thus, $q!/n!(q-n)!$ times degenerate.
The above spectra will come handy later.

\subsection{Asymptotic Bethe Ansatz approach}

We already mentioned that there are elaborate algebraic approaches
to derive the spectrum of the spin-Sutherland model, based on the
Yangian symmetry. We will, instead, take a lower-key approach which
reproduces the same spectra and is physically more lucid, although
not as rigorous. We will take the ABA route.

Consider {\it distinguishable} particles of the exchange-Calogero 
type without external potential, coming in with asymptotic
momenta $k_i$ and scattering off each other. Before scattering,
their positions are in some definite ordering determined by 
the ordering of their momenta (it is the inverse of that ordering).

The key observation is that, after scattering, the particles have
simply `gone through' each other with {\it no} backscattering \cite{SS}.
The impenetrable $1/x^2$ potential has become completely
penetrable in the presence of the exchange term! You can prove
this fact by examining the asymptotic properties of a simultaneous
eigenstate of $\pi_1 , \dots \pi_N$ which is obviously an energy
eigenstate: at $x_i \to \pm \infty$ the prepotential terms are vanishing
and we simply have eigenstates of the individual $p_i$. Since there are 
{\it no} pieces with the values of $p_i$ permuted (coming from
backscattering) we have complete transmission. 
(To explicitly see how it works, it is instructive to consider the 
two-body problem, decompose it into symmetric and antisymmetric parts,
scatter and recombine the parts after scattering. A relative
phase of $\pi$ between the two parts is what produces the effect.)

(Puzzle: what happens with the correspondence principle? 
With $\hbar$ back in, the interaction coefficient
is $\ell (\ell - \hbar M_{ij} )$. How can a term of order $\hbar$
produce such a dramatic effect, particles going through each
other, in the $\hbar \to 0$ limit?)

So the only effect of the scattering is a phase shift
of the wavefunction which, as we have said, is the sum of two-body
phases
\be
\theta_{sc} = \frac{N(N-1)}{2} \pi \ell
\ee
This is true on an infinite space. On a periodic space we can still
use the above result, together with the requirement for periodicity for
the wavefunction, to derive the spectrum. This is the ABA method and
is expected to reproduce the correct results in the thermodynamic limit
of many particles at constant density \cite{SS}.
It gives, in fact, the {\it exact} answer for the Sutherland model
\cite{Suth}, so we can expect it to work also in the present case.
For a space of period $2\pi$ the result is
\be
2\pi k_i + \sum_\pi \ell sgn(k_i - k_j ) = 2\pi n_i
\ee
The left hand side counts the total
phase picked up by a particle going round the space and scattering off 
to the other particles in the way. $n_i$ are arbitrary integers, ensuring
periodicity. There are, however, some constraints on the choice
of $n_i$ that are imposed by continuity from the $\ell =0$ case:

\noindent
--If $k_i \le k_j$ then $n_i \le n_j$

\noindent
--If $n_i = n_j$ there is a {\it unique} solution, that is,
$k_i < k_j$ and $k_i > k_j$ represent the same state.

\noindent
These rules are important to avoid overcounting and to discard
spurious solutions. With these, the spectrum obtained is the same
as the one derived with more rigorous methods. For the ordering
$n_1 \le \dots n_N$ the solution for $k_i$ is
\be
k_i = n_i + \ell (i-\frac{N+1}{2} )
\label{ABAk}
\ee
and similarly for other orderings. We see that the ABA momenta $k_i$ are
the same as the pseudomomenta that we have previously defined.

The bottom line is that the spectrum and degeneracies are the
same as those of distinguishable particles obeying generalized 
selection rules for their momentum. Still, what fixes the degeneracy
of states is the different ways that we can distribute the particles
to the quantum numbers $n_i$, rather than $k_i$ (see, especially,
the second rule above). A state of $N$ particles with the same
$n_i$, for instance, is nondegenerate although they have different
$k_i$ which would seemingly imply a permutation degeneracy.

For particles with spin the construction above, in combination with
the trick of the previous subsection of starting with fermions or
bosons, produces a spectrum with degeneracies the same as those
of free particles (the $n_i$ are `free' quantum numbers). As argued
before, for ferromagnetic interactions we must choose bosons and
combine their spins accordingly, while for antiferromagnetic 
interactions we must choose fermions. To spell it out, this means
the following:

\noindent
1. Choose a set of quantum numbers $n_i$. The ordering is immaterial,
since we have identical particles, so you can choose $n_1 \le \dots n_N$.

\noindent
2. Place your particles on these quantum numbers and put their
spins in the appropriate state. For the ferromagnetic case treat
them as bosons: the total spin of particles with the same $n_i$ transforms
in the symmetric tensor product of their spins. For the antiferromagnetic
case treat them as fermions: the total spin of particles with the
same $n_i$ transforms in the antisymmetric tensor product of their
spins; clearly up to $q$ can have the same $n_i$ in this case.

\noindent
3. Calculate the energy of this state in terms of the ABA momenta
(\ref{ABAk}): $E = \sum_i k_i^2$.

\noindent
It should be obvious that similar rules applied to the spin-Calogero
system reproduce the spectrum derived in the last subsection.
This method can be used to calculate both the statistical mechanics
(large $N$) of these systems and the few-body spectra.

\subsection{The freezing trick and spin models}

Now that we have a tractable way of solving spin-Calogero systems with
arbitrary strength of interaction we can introduce the freezing
trick \cite{PolE} and deal with spin chain models.

Consider, first, the previous ferromagnetic or antiferromagnetic
spin-Sutherland model. Take the limit $\ell \to \infty$.
The potential between the particles goes to infinity, so for
any finite-energy state the particles will be nearly `frozen' to 
their classical equilibrium positions. In fact, even the excitation
energies around that configuration will go to infinity:
the ground state energy scales like $\ell^2$, while the excitations
scale like $N \ell n + n^2$ with $n$ some excitation parameter.
So, to leading order in $\ell$ the spectrum becomes linear and of order 
$\ell$. These excitations correspond, essentially, to phonon modes
of small oscillations around the equilibrium positions of particles. The
`stiffness' of oscillations is, of course, proportional to the 
strength of the potential $\ell^2$ and the spectrum is proportional
to the frequency, of order $\ell$.

The quantum fluctuations of the particle positions in any state 
will scale like the inverse square root of the oscillator frequency,
that is, like $1/\sqrt \ell$. But, in the hamiltonian,
the piece coupling the spins to the kinematical degrees of freedom
is proportional to $1/ \sin ^2 x_{ij}$. In the large-$\ell$ limit, thus,
this term becomes a constant equal to its classical equilibrium value; 
so, in that limit, spin and kinematical degrees of freedom
decouple. (Note that the spin part is also of order $\ell$.)
The hamiltonian becomes
\be
H = H_S + \ell H_{spin}
\ee
with $H_S$ the spinless Sutherland hamiltonian and $H_{spin}$
the spin part
\be
H_{spin} = \mp \sum_{i<j} \frac{ 2 {\vec S}_i \cdot {\vec S}_j}
{4\sin^2 \frac{{\bar x}_{ij}}{2}}
\label{Hsp}
\ee
where the classical equilibrium positions ${\bar x}_j$ are equidistant
points on the circle:
\be
{\bar x}_j = \frac{2\pi j}{N}
\ee

The hamiltonian (\ref{Hsp}) above describes a spin chain consisting
of a regular
periodic lattice of spins in the fundamental of $SU(q)$ coupled
through mutual ferro- or antiferromagnetic interactions of strength
inversely proportional to their chord distance. It is the well known
$SU(q)$ Haldane-Shastry (HS) model \cite{Hal,Sha}. 
According to the above, its spectrum
can be found by taking the full spectrum of the corresponding
spin-Sutherland model in the large-$\ell$ limit, `modding out'
the spectrum of the spinless model and rescaling by a factor $1/\ell$.
Each state will inherit the spin representation of its `parent' 
spin-Sutherland state. So, both the energy and the total spin of
the states of the HS model can be determined this way. Commuting
integrals of this model \cite{Ino} can also be obtained this way
\cite{FM}. At the level of the partition function at some 
temperature $T$ we have
\be
Z_{spin} (T) = \lim_{\ell \to \infty} \frac{Z (\ell T)}
{Z_S (\ell T)}
\ee
From this, the thermodynamics of the spin chain model can be 
extracted.

We will not give the details of this construction here. We urge
anyone interested to solve this way a few-site (two or three) spin
chain, see how it works and deduce the `construction rules' for the
spectrum of a general spin chain. Let us
simply state that the many degeneracies of the spectrum of the
HS model (larger that the total spin $SU(q)$ symmetry
would imply), which is algebraically explained by the existence of
the Yangian symmetry, can, in this approach, be explained in terms
of the degeneracies of free particles. (The degeneracies are not 
{\it identical}, due to the modding procedure, but related.)

For the spin-Calogero model a similar limit can be taken, scaling also
the external oscillator frequency as $\omega \to \ell \omega$ to keep
the system bound. The classical equilibrium positions of 
this model are
at the roots of the $N$-th Hermite polynomial. We obtain, therefore,
a non-regular lattice of spins interacting with a strength inversely
proportional to the square of their distance \cite{PolE}. The spectrum
of this model can be found quite easily with the above method. Again,
we refer to the literature for details \cite{PolD,Fra,MX}.

In the continuum limit ($N \to \infty$) the antiferromagnetic version
of both the above models become $c=1$ conformal field theories, the HS
containing both chiral sectors while the inhomogeneous harmonic
one containing just one sector.

Other models exist and can be solved in this spirit: hierarchical
(many-coupling) models \cite{KwK}, supersymmetric models \cite{KtK,HBUW},
`twisted' models \cite{FK} etc. 
All, however, work only for the fundamental representation of some
internal group. The big, important open problem is to crack a particle
system with a {\it higher} representation for the spins and 
{\it arbitrary} coupling strength. If this is done, through the
freezing trick we will be able to solve a spin chain with spins
in a higher representation. This is interesting since we could then
see if the antiferromagnetic system for integer $SU(2)$ spins develops
a mass gap, according to the Haldane conjecture \cite{HG}.

\section{Folding the Calogero model: new systems by reduction}

In previous sections we presented mainly two variants of the Calogero
model: the rational (Calogero proper) and the trigonometric (Sutherland)
one. There are other possibilities, in which the two-body potential is
further generalized to a Weierstrass function (and its hyperbolic reduction)
as well as systems with `reflection' symmetry. Spin-Calogero models along
these generalizations and involving spin `twists' are also possible.

Rather than giving an independent analysis of these models, we prefer to
present how they can be obtained as appropriate reductions of standard
(spin) Calogero models. The advantage is a conceptual unifiation of the
various models, and the realization that the `root' model of particles with
inverse-square interaction contains all the fundamental information about
these systems, everything else stemming from it as particular sectors and
reductions.

\subsection{Reduction of spinless models}

The main idea is to reduce the
Calogero model by some of its discrete symmetries, akin to the `orbifold'
construction in field and string theory.

Consider any hamiltonian dynamical sysstem with some discrete symmetries $D$.
Its equations of motion remain invariant under the phase space mapping 
$\phi \to D(\phi)$, where $\phi$ are phase space variables. Then the
reduction to the invariant subspace $\phi = D(\phi)$ is kinematically
preserved; that is, the equations of motion do not move the system out
of this subspace. Therefore, reducing the initial value data to this
subspace trivially produces a system as solvable as the original one.
The motion will be generated by the original hamiltonian on the 
reduced space.

The starting point will be the plain vanilla scattering Calogero model
\be
H = \sum_{i=1}^N \half p_i^2 + \half \sum_{i \neq j} 
{g \over x_{ij}^2}
\ee
where $x_{ij} = x_i - x_j$.  The considered symmetries are:

$\bullet$ Translation invariance $T$: $x_i \to x_i + \ba$, $p_i \to p_i$

$\bullet$ Parity $P$: $x_i \to - x_i$, $p_i \to - p_i$

$\bullet$ Permutation symmetry $M$: $x_i \to x_{M(i)}$, $p_i \to p_{M(i)}$
with $M$ any element of the permutation group of $N$ particles.

Other symmetries will not be useful for our purposes.

A direct reduction of the system by any of the above symmetries
does not produce anything nontrivial or sensible: $\phi = T(\phi)$
is possible only in the trivial case $a=0$, while $\phi = P(\phi)$
and $\phi = M(\phi)$ requires (some) of the particle coordinates to
coincide, which is excluded by the infinite two-body potential.
We get useful systems only when reducing through appropriate products
of the above symmetries. These are:

\vskip 0.2cm

a) $D=PM$: We reduce by $P$ and a particular permutation:
$M(i) = N-i+1$
(or any other in the same conjugacy class). $M$ is uniquely fixed
from the fact that $D^2 (\phi) = M^2 (\phi)$ is a pure permutation and
for it not to make any two different particle coordinates to coincide
we must have $M^2=1$. Further, if for any particle $M(i)=i$ then
$D(x_i ) = -x_i$ and the corresponding coordinate is set to zero.
So $M(i) = i$ can happen for at most one $i$ (so that no two or
more particle coordinates are put to zero). So $M$ must be a collection
of rank two (two-body) permutations, and possibly a rank one (trivial)
permutation if the number of particles is odd, which can always be
represented in the form above.

The constraint
\be
x_i = -x_{N-i+1} ~,~~~p_i = -p_{N-i+1}
\ee
effectively reduces the 
original system into two mirror-images. The reduced hamiltonian is
\begin{eqnarray}
H &=& \sum_{i=1}^{N'} \half p_i^2 
+ \half \sum_{i \neq j} {g \over (x_i - x_j )^2}
+ \half \sum_{i \neq j} {g \over (x_i + x_j )^2}
+ \sum_i {g' \over x_i^2} \nonumber \\
N' &=& \left[\frac{N}{2} \right] ~,~~~ g' = g \left(\frac{1}{4} 
+ 2\left\{ \frac{N}{2} \right\} \right)
\end{eqnarray}
where an overall factor of $2$ has been discarded and $[.]$, $\{ . \}$
denote integer and fractional part, respectively.
The second term in the potential is the interaction
of each particle with the mirror image of each other particle; the
third part accounts for the interaction of each particle with the
mirror image of itself, and with a particle fixed at the origin by the
constraint (for odd $N$).

Parity symmetry persists in the case where 
an external harmonic oscillator potential is added to the system,
promoting it to the confining, rather than scattering, Calogero model:
\be
H = \sum_{i=1}^N \half p_i^2 + \half \sum_{i \neq j} 
{g \over x_{ij}^2} + \sum_{i=1}^N \half \omega^2 x_i^2
\ee
Reduction by $D=PM$ produces an integrable system similar as above with
the added harmonic oscillator potential.

\vskip 0.2cm

b) $D=TM$: No finite-rank element of the permutation group will do, 
since repeated application of $D$ would eventually lead to $x_i = x_i +m\ba$.
We overcome this by starting with $N'$ particles and taking the limit 
$N' \to \infty$. We pick the element: $M(i) = i+N$ for some finite $N$ 
which, for infinite $N'$, is infinite-rank. The constraint: 
\be
x_{i+N} = x_i +\ba ~,~~~ p_{i+N} = p_i
\ee
leads to a system consisting of infinitely many
copies of a finite system displaced by multiples of $\ba$. 
We can parametrize the particle indices by the pair $(i,m)$,
$i =1 \dots N$ and $m \in {\bf Z}$, where the original index
is $i+mN$. The constraint now reads
\be
x_{i,m} = x_i +m\ba ~,~~~ p_{i,m} = p_i
\label{conb}
\ee
The resulting system is infinite copies of an $N$-body system.
The reduced hamiltonian is
\begin{eqnarray}
H &=& \sum_{m=-\infty}^\infty \sum_{i=1}^N  \half p_i^2 + 
\half \sum_{m,n=-\infty}^\infty \sum_{i,j} 
{g \over (x_{ij} +m\ba -n\ba)^2}
\nonumber \\
&=& \half \sum_{m'=-\infty}^\infty \left\{\sum_{i=1}^N p_i^2 + 
\sum_{m=-\infty}^\infty \sum_{i \neq j} 
{g \over (x_{ij} +ma )^2} + \sum_i \sum_{m \neq 0}
{g \over (ma)^2} \right\}
\end{eqnarray}
In the above summation terms with ($i=j$, $m=n$) are omitted since
they correspond to self-interactions
of particles that are excluded from the original Calogero model.
The summation over $m'$ above accounts for the infinite periodically
repeating copies of the system and can be dropped. 
The infinite $m$-summation accounts for the interaction of each 
particle with the multiple images of each other particle and can
be performed explicitly. We eventually get
\be
H = \sum_{i=1}^N \half p_i^2 + \half 
\sum_{i \neq j} {g \pi^2 \over \ba^2 \sin^2 \pi {x_{ij} \over \ba}}
\ee
that is, the Sutherland model. 
In the above we omitted an irrelevant constant term equal to 
$gN {\pi^2 \over 6\ba^2}$, arising from the last term in the summand
corresponding to particle terms with $i=j$, which account
for the interaction of each particle with its own infinite images.

\vskip 0.2cm

c) We can formally extend the $T$ symmetry to complex parameter $\ba$.
As long as there is a subset of coordinates in the reduced phase
space that remains real and generates all other coordinates through
use of $D$, we will have a well-defined real subsystem. Applying
$D=TM$ for infinitely many particles parametrized by a double index
$(i,j)$, $i,j \in {\bf Z}$, for two complex translations $\ba$ and $\bb$,
the constraint is 
\be
x_{i,j} = x_{i+N',j} +\ba = x_{i,j+M'} +\bb ~,~~~
p_{i,j} = p_{i+N',j} = p_{i,j+M'} 
\ee
and we end up with a finite system with $N = M'N'$ particles periodically
repeating on the complex plane. (Clearly the specific separation of $N$ into
$N'$ and $M'$ is irrelevant; we could take, e.g., $N' = N$, $M' = 1$.)
Similarly to the Sutherland case,
the hamiltonian becomes infinitely many copies of
\be
H = \sum_{i=1}^{N'} \half p_i^2 + \half \sum_{i,j} 
\sum_{m,n=-\infty}^\infty {g \over (x_{ij} +m\ba+n\bb)^2}
\ee
The above sum has a logarithmic ambiguity that is easily regulated
by subtracting the constant $g/(m\ba+n\bb)^2$ from each term.
We end up with
\be
H = \sum_{i=1}^{N'} \half p_i^2 
+ \half \sum_{i \neq j} g {\cal P}(x_{ij} |\ba,\bb) 
\ee
that is, s model with an elliptic Weierstrass potential.

\vskip 0.2cm

d) $D_1=PM$ together with $D_2=TM$: 
This is a combination of (a) and (b) above. We work
again with an infinite number of particles. We impose two constraints:
\begin{eqnarray}
x_{-j+1-\epsilon} = -x_j ~&,&~~~ p_{-j+1-\epsilon} = -p_j
\nonumber \\
x_{j+N'} = x_j +\ba ~&,&~~~ p_{j+N'} = p_j
\end{eqnarray}
where $\epsilon = 0,1$ (any other choice of $\epsilon$ is equivalent
to one of these).
Parametrizing $j=i+mN'$ by the pair $i,m$, $i=1 \dots N'$, $m \in 
{\bf Z}$, we have
\begin{eqnarray}
x_{N-i+1-\epsilon} = \ba-x_{i,m} ~&,&~~~ p_{N-i+1-\epsilon} = -p_i
\nonumber \\
x_{i,m} = x_i +m\ba ~&,&~~~ p_{i,m} = p_i
\end{eqnarray}
so we end up with a finite system of $N=[(N' -\epsilon)/2]$ particles. The 
reduced hamiltonian is infinitely many copies of
\begin{eqnarray}
H = \sum_{i=1}^{N} \half p_i^2 &+& \half 
\sum_{i \neq j} {g \pi^2 \over \ba^2 \sin^2 \pi {x_i - x_j \over \ba}}
+ \half 
\sum_{i \neq j} {g \pi^2 \over \ba^2 \sin^2 \pi {x_i + x_j \over \ba}}
\nonumber \\
&+& \half 
\sum_{i} {g' \pi^2 \over \ba^2 \sin^2 \pi {2x_i \over \ba}}
+ \half 
\sum_{i} {g'' \pi^2 \over \ba^2 \sin^2 \pi {x_i \over \ba}}
\end{eqnarray}
with
\be
g' = g \left( \frac{1}{2} + 8\left\{ \frac{N-\epsilon}{2} \right\}
\right) ~,~~~
g'' = g \left( \epsilon - 2\left\{ \frac{N-\epsilon}{2} \right\}
\right)
\label{gg}
\ee
The coordinates $x_i$ can all be taken in the interval $(0,\ba/2)$
and the particles interact with their infinite mirror-images with
respect to mirrors placed at $x=0$ and $x=\ba/2$ and with
particles fixed at $x=0$ (if $\epsilon=1$) and at $x=\ba/2$
(if $2\{ (N-\epsilon)/2 \}=1$).
A similar construction can be performed with two complex translations,
as in (c) plus one parity reversal. We obtain a similar model
but with elliptic functions appearing instead of inverse sine squares.

\subsection{Integrals of motion via reduction}

The above exhausts the possibilities for
spinless particles. Before we proceed to the more 
interesting case of particles with spin, it is instructive to
demonstrate how the above construction reproduces the 
conserved integrals of motion of the reduced system. We will consider 
case (b), as the most generic, case (a) being rather trivial.

The Lax matrix of the original scattering Calogero model, as given
in an early section, is
\be
L_{ij} = p_i \delta_{ij} + (1-\delta_{ij} ) \frac{i\ell}
{x_{ij}}
\ee
Traces of powers of $L$ produce the integrals of motion in 
involution for the model:
\be
I_k = \tr L^k ~,~~~k=1,\dots N
\ee

For the system of case (b), we promote the index $i$ into a
pair $(i,m)$ and choose $x_{i,m} = x_i + m\ba$ as in (\ref{conb}).
The resulting infinite-dimensional matrix $L_{im,jn}$ can be thought of
as consisting of infinitely many blocks of size $N \times N$,
$m,n$ labeling the blocks and $i,j$ the elements of each block:
\be
L_{im,jn} = p_i \delta_{mn} \delta_{ij}
+ (1-\delta_{mn} \delta_{ij} ) \frac{i\ell}{x_{ij} + (m-n)\ba}
\ee
We observe a block `translational invariance' of the matrix $L$ in
the indices $m,n$, which reflects the invariance of the model under 
a translation by $\ba$. Due to this, we can trade the pair
$m,n$ for a single index $m-n$ 
\be
L_{im,jn} \equiv L_{m-n;ij}
\ee
and thus $L$ becomes an infinite
collection of $N \times N$ matrices $L_n$ labeled by $n$.
We define the Fourier transform $L(\sigma )$:
\be
L(\sigma) = \sum_n e^{in\sigma} L_n 
\ee
in terms of which $L_n$ is
\be
L_n = \frac{1}{2\pi} \int_0^{2\pi} L(\sigma) e^{-in\sigma}
\ee
The corresponding integrals of motion $I_k$ are traces of powers
of $L$. Denoting by $\Tr$ the trace in the infinite-dimensional
space labeled by $i,m$ and by $\tr$ the trace in the $N$-dimensional
space labeled by $i$ alone, we have:
\begin{eqnarray}
I_k = \Tr L^k &=& \sum_{n_1 , \dots n_k} \tr( L_{n_1 -n_2} 
\cdots L_{n_k -n_1} ) \nonumber \\
(m_i \equiv n_i - n_{i+1})~~~
&=& \sum_{n_1} \, \sum_{m_1 , \dots m_{k-1}}
\tr ( L_{m_1} \cdots L_{m_{k-1}} L_{-m_1 \cdots -m_{k-1}} )
\end{eqnarray}
The sum over $n_1$ above produces a trivial infinity. This is
due to the summation over the infinite copies of the system,
just as in the case of the hamiltonian, and will be dropped.
In terms of the Fourier transformed $L(\sigma)$ the reduced
$I_k$ become simply
\be
I_k = \frac{1}{2\pi} \int_0^{2\pi} \tr L(\sigma)^k d\sigma
\label{Ik}
\ee
It is now a matter of calculating $L(\sigma)$. From the
Fourier transform
\be
\sum_n \frac{e^{in\sigma}}{n+x} = \frac{\pi e^{i(\pi -\sigma)x}}
{\sin \pi a} ~~~{\rm for}~~0<\sigma<2\pi
\label{sum}
\ee
we obtain for $L(\sigma)$
\be
L(\sigma)_{ij} = e^{i(\pi-\sigma)x_{ij}} \left[
p_i \delta_{ij} + (1-\delta_{ij})\frac{i\pi\ell}{\ba \sin\pi
\frac{x_{ij}}{\ba}} -\ell\frac{\pi-\sigma}{\ba}\delta_{ij} \right]
\label{Ls}
\ee
where the last diagonal term linear in $\pi-\sigma$ came from
terms with $i=j$, $n \neq 0$ in $L_{n,ij}$. The
matrix inside the square bracket, apart from this linear part,
is the Lax matrix $\tilde L$ of the Sutherland model:
\be
{\tilde L}_{ij} = p_i \delta_{ij} + (1-\delta_{ij})
\frac{i\pi\ell}{\ba \sin\pi\frac{x_{ij}}{\ba}} 
\ee

Substituting (\ref{Ls}) in (\ref{Ik}) we note that the
exponential factors cancel (due to $x_{i_1 i_2} + \cdots x_{i_{k-1}
i_1} =0$) and we are left with
\begin{eqnarray}
I_k &=& \frac{1}{2\pi} \int_0^{2\pi} \tr \left( {\tilde L} -\ell \frac{
\pi-\sigma}{\ba} \right)^k d\sigma
= \sum_{s=0}^k \tr \frac{k!}{s! (k-s)!} {\tilde L}^{k-s} 
\frac{(-\ell)^s}{2\pi \ba^s} 
\int_0^{2\pi} (\pi-\sigma)^s d\sigma \nonumber \\
&=& \sum_{n=0}^{[k/2]} 
\frac{k!}{(2n+1)! (k-2n)!} 
\left( \frac{\pi \ell}{\ba} \right)^{2n}
~ {\tilde I}_{k-2n} 
\end{eqnarray}
where ${\tilde I}_k = \tr {\tilde L}^k$ are the conserved integrals
of the Sutherland model. We obtain a linear combination
of the intergal ${\tilde I}_k$ and lower integrals of the same parity.
The appearence of the lower integrals originates from the interaction
of each particle with its own infinite images. We saw an example of
such a term in the constant potential term that we omitted from the
reduced hamiltonian of case (b). In conclusion, we have 
recovered the integrals of the Sutherland model.

\subsection{Reduction of spin-Calogero systems}

We extend now these considerations to systems of particles with internal 
classical $U(n)$ degrees of freedom. The 
corresponding starting spin-Calogero system can be obtained by nontrivial
reductions of the hermitian matrix model, analogous to the ones that led
to the spin-Stherland model from the untary matrix model. Alternatively,
we may otain the spin-Calogero model by taking the infinite-volume limit
of the periodic spin-Sutherland model. We shall also consider the classical
version of the model, in which the spins become classical phase space
variables whose Poisson brackets generate the $U(n)$ algebra.

The hamiltonian of this model reads
\be
H = \sum_{i=1}^N \half p_i^2 + \half \sum_{i \neq j} {\tr(S_i S_j )
\over x_{ij}^2}
\ee
The $S_i$ are a set of independent classical $U(n)$ spins of rank one 
and length
$\ell$, that is, $n \times n$ rank-one hermitian matrices satisfying
\be
\tr (S_i )^2 =  \ell^2
\label{lengthl}
\ee
and with Poisson brackets
\be 
\{ (S_i)_{ab} , (S_j)_{cd} \} = -i \delta_{ij}
\left[ (S_i)_{ad} \delta_{cb} - \delta_{ad} (S_i)_{cb} \right]
\ee
This can be thought of as the classical limit of the spin models
obtained in the quantum treatment of the matrix model. The spin
representation was restricted to be totally symmetric, which
classically translates to the rank-one condition.

Just as in the quantum case, such spins can be realized in terms of
classical oscillators: 
\be
(S_i )_{ab} = {\bar A}_i^a A_i^b
~,~~ a,b=1 \dots n
\label{Sa}
\ee
where $( A_i^a \,,\, {\bar A}_i^a)$ are a set of $nN$ independent
classical harmonic oscillator canonical pairs with Poisson brackets:
\be
\{ A_i^a , {\bar A}_j^b \} = i \delta_{ij} \, \delta_{ab}
\ee
and satisfying the constraint
\be
\sum_a {\bar A}_i^a A_i^a = \ell ~~{\rm for~all}~i
\label{Aconc}
\ee

The above model, in addition to the previous symmetries $T$, $P$ and
$M$, also possesses the symmetry

$\bullet$ Spin rotations $U$: $S_i \to U S_i U^{-1}$, with $U$ a 
constant unitary $n \times n$ matrix. $(x_i , p_i )$ remain unchanged.

Again, reduction by this symmetry alone leads to no interesting
system (implying either $U=1$ or $S_i=0$). Reduction by $PUM$
or $TUM$, however, much along the lines of the previous $PM$ 
and $TM$ reductions, produces new and nontrivial results:

\vskip 0.2cm

e) $D=PUM$ with $P$ and $M$ as in (a) before, and $U$ a unitary
matrix satisfying $U^2 =1$ (this is necessary since $P$ and $M$
are of rank two). The constraints are
\be
x_i = -x_{N-i+1} ~,~~~p_i = -p_{N-i+1} ~,~~~ S_i = U S_{N+i-1} U^{-1}
\ee
The reduced hamiltonian acquires the form:
\begin{eqnarray}
H = \sum_{i=1}^{N'} \half p_i^2 
&+& \half \sum_{i \neq j} {\tr(S_i S_j) \over (x_i - x_j )^2}
+ \half \sum_{i \neq j} {\tr(S_i U S_j U^{-1}) \over (x_i + x_j )^2}
\nonumber \\
&+& \sum_i {\tr(S_i U S_i U^{-1}) \over 4x_i^2}
+ g\sum_i {\tr(S_i S_o) \over x_i^2}
\end{eqnarray}
where $N' = [N/2]$, $g=2\{N/2\}$ and $S_o = U S_o U^{-1}$ 
is an extra spin degree of freedom.

The form of the hamiltonian for the reduced model and its physical
interpretation simplifies with an appropriate choice of basis for
the spins: by using the $U$-invariance of the full model, we can
perform a unitary rotation $V$ to all spins $S_i \to V S_i V^{-1}$.
This transforms the matrix $U$ appearing in (\ref{TUMd}) into
$U \to V^{-1} U V$. With an appropriate choice of $V$ we can always
choose $U$ to be diagonal: $U = diag (e^{i \phi_a})$. Because of
the constraint $U^2 =1$, this means that $U$ can take the form
$U = diag(1,\dots 1,-1,\dots -1)$ with $n_1$ ($n_2$) entries equal
to $1$ (resp.~$-1$). So we see that the original $U(n)$ invariance
of the model has been broken to $U(n_1) \times U(n_2)$. If $n_1
= n_2$ there is an additional $Z_2$ exchange symmetry.

As in the spinless case (a), we could have started with a spin-Calogero
model in an external oscillator potential (which shares the same
$U$ and $P$ symmetries), and obtain a model as above with the
extra confining harmonic potential.

\vskip 0.2cm

f) $D=TUM$ with $T$ and $M$ as in (b) before, and $U$ any unitary
matrix. The constraint on the phase space is
\be
x_i = x_{i+N} +\ba ~,~~~ p_i = p_{i+N} ~,~~~ S_{i+N} = U S_i U^{-1}
\label{TUMd}
\ee
The system becomes, again, infinite copies of $a$-translated and
$U$-rotated systems, and the reduced hamiltonian is
\be
H = \sum_{i=1}^N \half p_i^2 + \half \sum_{i,j} 
\sum_{m=-\infty}^\infty {\tr(S_i U^m S_j U^{-m} ) 
\over (x_{ij} +m\ba )^2}
\label{HTUMd}
\ee
In the above we cannot drop terms with $i=j$ any more, since they are
now spin-dependent rather than constant. Only the term ($i=j$, $m=0$)
must be dropped from the summation as before.

Again, the form of the hamiltonian for the reduced model and its physical
interpretation simplifies with an appropriate choice of basis for
the spins which makes $U$ diagonal: $U = diag (e^{i \phi_a})$. 
The trace in (\ref{HTUMd}) then becomes
\be
\tr ( S_i U^m S_j U^{-m} ) = \sum_{a,b=1}^n (S_i)_{ab} (S_j)_{ba}
e^{-im \phi_{ab}}
\ee
where $\phi_{ab} = \phi_a - \phi_b$. The 
$m$-summation appearing in (\ref{HTUMd}) gives
\begin{eqnarray}
\sum_{m=-\infty}^\infty {\tr(S_i U^m S_j U^{-m} )\over 
(x_{ij} +m\ba )^2} &=&
\sum_{a,b=1}^n (S_i)_{ab} (S_j)_{ba} \sum_{m=-\infty}^\infty
{e^{-im \phi_{ab}} \over (x_{ij} +m\ba )^2 } \\
&=& \sum_{a,b=1}^n V_{ab} (x_{ij}) \, (S_i)_{ab} (S_j)_{ba} 
\end{eqnarray}
with the potential $V_{ab} (x)$ being
\be
V_{ab} (x) = \sum_{m=-\infty}^\infty
{e^{-im \phi_{ab}} \over (x +m\ba )^2 } 
\ee
We must distinguish between the cases $i \neq j$ and $i=j$.
For the case $i \neq j$ the sum can be obtained from the
$x$-derivative of (\ref{sum}):
\be
V_{ab} (x) = \frac{1}{\ba^2} e^{-i{x \over \ba}\phi_{ab}}
\left( {\pi^2 \over \sin^2{\pi x \over \ba}} -i\pi \phi_{ab}
\cot{\pi x \over \ba} - \pi | \phi_{ab} | \right)
\ee
For the case $i=j$ we must omit the term $m=0$ from the summation.
We obtain an $x$-independent potential:
\be
{\tilde V}_{ab} \equiv \lim_{x \to 0} \left( V_{ab} (x) -
\frac{1}{x^2} \right) = {\phi_{ab}^2 \over 2\ba^2} - 
{\pi | \phi_{ab} | \over \ba^2}
\ee
In the above we omitted a constant ($a,b$-independent) term equal
to $\frac{\pi^2}{3\ba^2}$ which would contribute to the hamiltonian
a term proportional to $\sum_i \tr (S_i )^2$. Due to (\ref{lengthl}), 
this is an irrelevant constant. 
With the above, the reduced hamiltonian eventually becomes 
\be
H = \sum_{i=1}^N \half p_i^2 + \half \sum_{i \neq j} 
\sum_{a,b} V_{ab} (x_{ij}) \, (S_i)_{ab} (S_j)_{ba} +
\half \sum_i \sum_{a,b} {\tilde V}_{ab} \, (S_i)_{ab} (S_i)_{ba} 
\label{Hab}
\ee
This is a model of particles with $U(n)$ spins interacting through
$U(n)$ {\it non}-invariant couplings, due to the presence of the
matrix $V_{ab}$. The original global $U(n)$ invariance is, now, broken
to the diagonal $U(1)^n$ part and only the diagonal components $S_{aa}$
of the total spin
\be
S_{ab} = \sum_i (S_i)_{ab}
\ee
are conserved.
The standard $U(n)$-invariant spin-Sutherland model is recovered
upon choosing $V_{ab} \sim \delta_{ab}$, in which case the sums
over $a,b$ above become a normal trace. This is achieved by choosing
$\phi_a$=constant, that is, $U=e^{i\phi}$.

The above model is, in fact, the same as the classical model
introduced by Blom and Langmann \cite{BL}, and the author of this review
\cite{AP}, in the particle-spin form in which it was recast
in \cite{AP}:
\begin{eqnarray}
H &=& \half \sum_i p_i^2 
+ \half \sum_{i \neq j} \left( \sum_{ij} V_{ab} (x_{ij} ) 
({\hat S}_i)_{ab} ({\hat S}_j)_{ba} + {\ell (\ell +n) \over 4n 
\sin^2 {x_{ij} \over 2}}\right) \\
&+& \half \sum_i \sum_{ab} {\tilde V}_{ab} \, ({\hat S}_i)_{aa} 
({\hat S}_i)_{bb} + {1 \over 2N} \sum_{ab} {\tilde V}_{ab} \left( q_a q_b
- {\hat S}_{aa} {\hat S}_{bb} \right)
\label{Hspc}
\end{eqnarray}
To fully see the equivalence, we must observe
the following:

1. In the present construction we expressed the hamiltonian in terms
of $U(n)$ spins $S_i$. In \cite{AP} it was, instead, expressed in terms of
traceless $SU(n)$ spins ${\hat S}_i$. By (\ref{Sa}) and (\ref{Aconc}) we 
have $\tr S_i = \ell$, so the relation between the two is
\be
{\hat S}_i = S_i - \frac{\ell}{n}
\ee

2. The expression (\ref{Hspc}) derived in \cite{AP} was fully
quantum mechanical. It can be seen that the term $\ell (\ell +n)$
in (\ref{Hspc}) classically becomes $\ell^2$ ($n$ was a quantum
correction similar to the shift of the classical angular momentum
$J^2$ to $J(J+1)$).

3. For the rank-one matrices $S_i$ we have the relation
\be
(S_i)_{ab} (S_i)_{ba} = (S_i)_{aa} (S_i)_{bb}
\ee

4. In \cite{AP} a set of dynamically conserved charges $q_a$
were introduced that can be chosen to have any value as long
as they sum to zero.

5. In \cite{AP} the particles were taken to move on the unit
circle, that is, $\ba=2\pi$.

Doing the above substitutions in (\ref{Hab}) we see that it becomes
practically identical to (\ref{Hspc}). The two expressions  differ
by constant terms depending on the charges $q_a$
and the diagonal elements of the total spin $S_{aa}$. Since both
of these quantities are constants of the motion, the two models
are trivially related.

\vskip 0.2cm

g) We can, similarly to (c), extend the above construction to two
complex translations and corresponding spin rotations. Parametrizing
again the infinite number of particles with a doublet of indices
$i,j \in {\bf Z}$ the constraints are
\begin{eqnarray}
x_{i,j} = x_{i+N' ,j} +\ba ~,~~~ p_{i,j} &=& p_{i+N' ,j} 
~,~~~ S_{i+N' ,j} = U S_{ij} U^{-i} \nonumber \\
x_{i,j} = x_{i,j+M'} +\bb ~,~~~ p_{i,j} &=& p_{i,j+M'} 
~,~~~ S_{i,j+M'} = V S_{i,j} V^{-1}
\end{eqnarray}
and we end up as before with a finite system with $N = M' N'$ 
particles periodically repeating on the complex plane.
Choosing $M' =1$, $N' =N$, the spin matrices 
$S_{i+mN,n} \equiv S_{i;m,n}$ are now expressed as
\be
S_{i;m,n} = U^m V^n S_i V^{-n} U^{-m} = 
V^n U^m S_i U^{-m} V^{-n}
\label{Ssol}
\ee
The corresponding reduced hamiltonian becomes infinite copies of
\be
H = \sum_{i=1}^{N} \half p_i^2 + \half \sum_{i,j} 
\sum_{m,n=-\infty}^\infty {\tr( U^m V^n S_i V^{-n} U^{-m} S_j ) 
\over (x_{ij} +m\ba +n\bb )^2}
\label{Hmn}
\ee

Since the
two space translations and the corresponding particle permutations
commute, for consistency the two spin rotations must also commute:
\be
UV S_{i;m,n} V^{-1} U^{-1} =
VU S_{i;m,n} U^{-1} V^{-1} 
\ee
which implies
\be
[ U^{-1} V^{-1} UV, S_{i;m,n} ] =0
\ee
For this to hold for generic $S_i$ we must require the `group commutator'
matrix
\be
\omega \equiv U^{-1} V^{-1} UV
\ee 
to be proportional to the identity matrix.
Clearly $\omega$ satisfies $\det(\omega) = \omega^\n =1$, so we obtain
\be
UV = \omega VU ~,~~~ \omega = e^{i2\pi\frac{\nu}{\n}}
\ee
with $\nu$ an integer $0\leq \nu <\n$. $U$ and $V$ then satisfy 
Weyl's braiding condition which characterizes a noncommutative 
(`quantum') torus \cite{Weyl}.

For $\nu=0$ ($\omega=1$) the matrices $U$ and $V$ commute. For $\omega \neq 1$,
however, $U$ and $V$ become `clock' and `shift' matrices. 
We shall deal with the two cases separately.

\subsection{Twisted elliptic spin-Calogero models}

In this case $U$ and $V$ commute and can be simultaneously diagonalized.
Just as in the case of a single $U$, we choose a basis for the spins
that diagonalizes both $U$ and $V$ to $U=diag( \phi_i )$,
$V=diag( \theta_i )$. The $m,n$-sums that appear in (\ref{Hmn})
become
\be
\sum_{a,b=1}^n (S_i)_{ab} (S_j)_{ba} 
\sum_{m,n=-\infty}^\infty { e^{-im\phi_{ab} -in\theta_{ab}}
\over (x_{ij} +m\ba+n\bb)^2}
\ee
where the term $m=n=0$ is omitted if $i=j$. 
We obtain again
a potential $V_{ab} ( x_{ij} )$, for $i \neq j$, given by the sum
\be
V_{ab} (x) =
\sum_{m,n=-\infty}^\infty { e^{-im\phi_{ab} -in\theta_{ab}}
\over (x +m\ba+n\bb)^2}
\ee
and a spin self-coupling ${\tilde V}_{ab}$ for $i=j$, given by
\be
{\tilde V}_{ab} = 
\sum_{(m,n) \neq (0,0)} { e^{-im\phi_{ab} -in\theta_{ab}}
\over (m\ba+n\bb)^2}
\ee
Note that now, due to the presence of the phase factors, these sums
are convergent and have no regularization ambiguity. The only
ambiguous terms, defined modulo an additive constant, are the ones with 
$a=b$. We will comment on the impact of such a regularization
ambiguity in the sequel.
%$V_{aa} (x)$ can be regularized in a way
%similar to the spinless case, yielding the Weierstrass potential,
%while we may choose ${\tilde V}_{aa}$ to vanish altogether.

The potential $V_{ab} (x)$ is a modular function on the complex torus 
$(\ba,\bb)$ with quasiperiodicity
\begin{eqnarray}
V_{ab} (x+\ba) &=& e^{i\phi_{ab}} \, V_{ab} (x) \nonumber \\
V_{ab} (x+\bb) &=& e^{i\theta_{ab}} \, V_{ab} (x) 
\end{eqnarray}
It has a double pole at $x=0$, with principal part
\be
V_{ab} (x) = \frac{1}{x^2} + O( x^0 )
\ee
and no other poles in each cell. These properties uniquely define
$V_{ab}$ and allow for an expression in terms of theta-functions.
We put
\be
V_{ab} (x) = A e^{i\frac{x}{\ba}\phi_{ab}} \frac{
\tha \left( \frac{\pi}{\ba} (x-q_1) \right)
\tha \left( \frac{\pi}{\ba} (x-q_2)\right)}
{\tha \left( \frac{\pi}{\ba} x \right)^2}
\label{VAqq}
\ee
where $q_{1,2}$ are the as yet unknown zeros of $V_{ab} (x)$
and the theta-functions appearing above have complex period
$T = \bb/\ba$. This has the right quasiperiodicity under
$x \to x+\ba$. In order to also have the right quasiperiodicity under
$x \to x+\bb$, $q_{1,2}$ must satisfy
\be
q_1 + q_2 = \frac{1}{2\pi} (\ba \theta_{ab} - \bb \phi_{ab} )
\equiv Q_{ab}
\label{qqQ}
\ee
and to have the right behavior around $x=0$ we must further have
\be
A = \frac{\pi^2 \tha' (0)^2}{\ba^2 
\tha \left( \frac{\pi q_1}{\ba} \right)
\tha \left( \frac{\pi q_2}{\ba} \right)} 
\label{Aqq} 
\ee
\be
\frac{ \tha' \left( \frac{\pi}{\ba} q_1 \right) }
{ \tha \left( \frac{\pi}{\ba} q_1 \right) } +
\frac{ \tha' \left( \frac{\pi}{\ba} q_2 \right) }
{ \tha \left( \frac{\pi}{\ba} q_2 \right) }
= i\frac{\phi_{ab}}{\pi}
\label{qqp}
\ee
The equations (\ref{qqQ}) and (\ref{qqp}) above determine
$q_1$ and $q_2$, while (\ref{Aqq}) then determines $A$. It may be
possible to express $q_1$, $q_2$ in a more explicit form, or to
recast (\ref{VAqq}) in a form more symmetric in $\ba$,$\bb$, by 
using theta-function identities. Finally, the self-coupling
${\tilde V}_{ab}$ can be extracted from $V_{ab} (x)$ as
\be
{\tilde V}_{ab} = \lim_{x \to 0} \left( V_{ab} (x) - \frac{1}{x^2}
\right)
\label{Vtil}
\ee

To sum up, we obtain a `twisted' $U(n)$ non-invariant spin-generalization
of the elliptic model given by a hamiltonian of the form (\ref{Hab}) but with
the potentials appearing now being given by (\ref{VAqq},\ref{Vtil}).
The $U(n)$ invariance of the original model is, again, broken
down to the diagonal abelian sungroup $U(1)^n$ due to the dependence of
the potential on $a,b$.
The $U(n)$-invariant spin-Weierstrass model is regained for $\phi_{ab}
= \theta_{ab} =0$, that is, trivial matrices $U$ and $V$. 

We point out that for $\theta_{ab} = \phi_{ab} =0$, that is,
$Q_{ab} =0$, the equations for $q_{1,2}$ (\ref{qqQ},\ref{qqp})
are satisfied for {\it any} $q_1 = -q_2$ leading to an apparent
arbitrariness. As can be seen, however, by applying the addition 
formula
\be
\tha (x+q) \tha (x-q) \thh (0)^2 =  \tha (x)^2 \thh (q)^2
- \thh (x)^2 \tha (q)^2
\ee
this simply amounts to an arbitrary additive constant to the
expression for $V_{ab} (x) \equiv V (x)$. This corresponds to the
need for regularization for this expression in the absence of
phases, as explained before.
(In the case of the Weierstrass function this is fixed by further
requiring that the $O( x^0 )$ part of the function at $x=0$ vanish,
which picks $q=\pi T/2$ and makes $\thh (q)$ above vanish.)
We also point out that we can pick {\it any} of these values
for $q_1 = -q_2$ at the limit $Q=0$ by appropriately choosing
the ratio $\phi_{ab} / \theta_{ab}$ as they both go to zero.

The ambiguity of the terms with $a=b$ can be fixed in the same
way: we can choose phases $\phi_{aa} \neq 0$, $\theta_{aa} \neq 0$,
evaluate the expressions, and then let $\phi_{aa} , \theta_{aa} \to 0$.
This will lead to arbitrary additive constants $C_a$, depending
on the ratio $\phi_{aa} / \theta_{aa}$ as we take them to zero.
The same constants, however, will appear in both $V_{aa} (x)$
and ${\tilde V}_{aa}$. Their net contribution to the hamiltonian
will be
\be
\Delta H = 
\half \sum_{i \neq j} \sum_a C_a \, (S_i)_{aa} (S_j)_{aa} +
\half \sum_i \sum_a C_a \, (S_i)_{aa} (S_i)_{aa} =
\half \sum_a C_a \, (S_{aa})^2
\ee
Since the diagonal components of the total spin $S$ are still
constants of the motion, due to the residual $U(1)^n$ invariance,
this amounts to the addition of an overall constant, and thus leads
to systems that are trivially related.

The same discussion applies if the angles $\phi_a$ and $\theta_a$
coincide for two or more values of $a$ belonging to a subspace
of indices $I$, in which case $\phi_{ab} = \theta_{ab} =0$ for 
$a,b \in I$. This will result to a
constant additive matrix $C_{ab}$ in the potential for this subspace
of indices $I$, leading to an extra contribution to the hamiltonian
\be
\Delta H = \sum_{a,b \in I} C_{ab} \, S_{ab} S_{ba}
\ee
Since $\phi_{ab} = \theta_{ab} =0$ for $a,b \in I$, however, the 
corresponding
subgroup of $U(n)$ remains unbroken, and thus the corresponding
components of the total spin $S_{ab}$ appearing above are constants
of the motion. Once again, the arbitrary terms are constant and we
essentially obtain a unique system.

Overall, this is a generalization of the spin-Weierstrass model
to one involving $2n$ phases that break the $U(n)$ invariance
and promote the potential to a modular function. The potential lives
on a complex torus in the coordinates, where translations around each 
nontrivial
cycle are accompanied by spin transformations. The model obtained
in (f) can be though of as the limit of the present model with
$\bb \to \infty$, in which case $\theta_a$ become
irrelevant.

\subsection{Noncommutative elliptic spin-Calogero model}

In this case
\be
UV = \omega VU ~,~~~ \omega = e^{i2\pi\frac{\nu}{\n}}
\label{UVncom}
\ee
with $\nu$ a nonvanishing integer $0 < \nu <\n$. 

To proceed, we must identify the form of $U,V$. We need the
irreducible representations of the relation (\ref{UVncom}). 
Call $k$ the greatest common divisor of $\nu$
and $\n$. Then $\n = k\m$ and $\nu = k\mu$, for relatively prime $\m,\mu$. 
The irreducible representations for $U,V$ are $\m$-dimensional
`clock' and `shift' matrices. By a global $U(\n)$ spin transformation
we can diagonalize either of $U,V$. Choosing $U$ diagonal, the general
form of $U$ and $V$ will be the direct sum of $k$ of the above irreducible
representations:
\be
U = diag\{e^{i\phi_0} , \dots e^{i\phi_{k-1}} \} \otimes u ~,~~~
V = diag\{e^{i\theta_0} , \dots e^{i\theta_{k-1}} \} \otimes v
\label{UVsol}
\ee
where $\phi_q, \theta_q$ are arbitrary phases, determining the
Casimirs $U^\m$ and $V^\m$, and $u,v$ are the $\m$-dimensional clock
and shift matrices
\be
u_{\alpha \beta} = \omega^{\alpha} \, \delta_{\alpha \beta} ~,~~~ 
v_{\alpha \beta} = \delta_{\alpha+1,\beta}~(mod~\m)
~,~~~ \alpha,\beta=0,\dots \m-1
\ee
So the acceptable $U$ and $V$ depend on $2k$ arbitrary parameters.

To take advantage of the form (\ref{UVsol}) for $U,V$ we partition
$S_i$ into $k^2$ blocks of dimension $\m \times \m$ each by using the
double index notation
\be
(S_i )_{ab} = (S_i )_{\alpha \beta}^{pq}~,~~~ 
a = p\m+\alpha~,~~b=q\m+\beta 
\ee
The $U(\n )$ Poisson brackets in this notation are
\be 
\{ (S_i )_{\alpha \beta}^{pq} , (S_j )_{\gamma \delta}^{rs} \} 
= -i \delta_{ij} \left[ 
(S_i)_{\alpha \delta}^{ps} \, \delta_{\gamma \beta} \, \delta_{rq} 
- \delta_{\alpha \delta} \, \delta_{rq} \, (S_i)_{\gamma \beta}^{rq} \right]
\label{SPB}
\ee
The $m,n$-sums that appear in (\ref{Hmn}) then 
become
\be
\sum_{m,n;\alpha,\beta;p,q} (S_i)_{\alpha+n,\beta+n}^{pq} \,
(S_j)_{\beta,\alpha}^{qp} \,
{e^{-im\phi_{pq} -in\theta_{pq}} \, \omega^{ m (\alpha-\beta)}
\over (x_{ij} +m\ba +n\bb)^2}
\label{pot}
\ee
where the term $m=n=0$ is omitted if $i=j$. 

The above gives a potential interaction between particles $i$ and $j$
in the form of a modular function in $x_{ij}$ which depends on the
spin components of particles $i$ and $j$. To make the noncommutative
character of the spin interaction manifest, we perform a change
of basis in the spin states and define
\be
({\tilde S}_i )_{\alpha \beta}^{pq} = \sum_\sigma \omega^{\left( \sigma 
+ \frac{\alpha}{2}\right)\beta} \, (S_i )_{\alpha+\sigma,\sigma}^{pq}
\label{Stil}
\ee
This is essentially a discrete Fourier transform in the sum
of the $\alpha,\beta$ indices of $S_{\alpha \beta}^{pq}$.
(Note that, for $\m$ odd, ${\tilde S}_{\alpha \beta}^{pq}$ is
actually antiperiodic in the index $\alpha$ if $\beta$ is odd,
and vice versa. Although we could have defined a properly periodic
matrix, we prefer this slight inconvenience in order to make the
ensuing formulae more symmetric.)
In fact, it will be convenient to assemble the double indices
$(\alpha,\beta)$ and $(m,n)$ into vectors $\vec \alpha$ and
$\vec m$. Similarly, we define $\vec \bc = (\ba , \bb )$
and ${\vec \phi}_p = (\phi_p , \theta_p )$.

The Poisson brackets of the ${\tilde S}_i$ are found from (\ref{SPB})
\be
\{ ({\tilde S}_i )_{\vec \alpha}^{pq} , 
({\tilde S}_j )_{\vec \beta}^{rs} \} = 
i \delta_{ij} \left[ \omega^{\frac{{\vec \alpha} \times {\vec \beta}}{2}}
\, \delta_{ps} \, ({\tilde S}_i )_{{\vec \alpha}+{\vec \beta}}^{rq}
- \omega^{-\frac{{\vec \alpha} \times {\vec \beta}}{2}} \,
\delta_{rq} \, ({\tilde S}_i )_{{\vec \alpha}+{\vec \beta}}^{ps}
\right]
\ee
This is a structure extending the Moyal (star-commutator) algebra, 
the exponent 
of $\omega$ being the cross product of the discrete `momenta' 
$\vec \alpha$ and $\vec \beta$. For $(rs)=(pq)$, in particular,
it becomes the torus Fourier transform of the Moyal bracket
\be
\{ ({\tilde S}_i )_{\vec \alpha}^{pq} , 
({\tilde S}_j )_{\vec \beta}^{pq} \}
= i \delta_{ij} (\omega^\half - \omega^{-\half} ) 
\left[ {\vec \alpha} \times {\vec \beta} \right]_\omega
({\tilde S}_i )_{{\vec \alpha}+{\vec \beta}}^{pq}
\ee
where 
\be
[x]_\omega = \frac{\omega^\frac{x}{2} - \omega^{-\frac{x}{2}}}
{\omega^\half - \omega^{-\half}}
\ee
is the $\omega$-deformation of $x$. This is the so-called
trigonometric algebra with periodic discrete indices \cite{FZ}.

Finally, by inverting (\ref{Stil}) and substituting in (\ref{pot}),
the potential energy $W$ in terms of the ${\tilde S}_i$ acquires the form
\be
W = \sum_{i,j} \sum_{{\vec \alpha};p,q} 
({\tilde S}_i )_{\vec \alpha}^{pq} ~ ({\tilde S}_j )_{-\vec \alpha}^{qp}
~ W_{\vec \alpha}^{pq} ( x_{ij} )
\label{Vfin}
\ee
The above includes two-body interactions, for $i\neq j$, as well as 
spin self-couplings, for $i=j$, arising from the interaction of each 
particle with its own images in different cells. The two-body potential 
$W_{\vec \alpha}^{pq} (x)$ is
\be
W_{\vec \alpha}^{pq} (x) = \frac{1}{\m} \sum_{\vec m}
\frac{ \omega^{{\vec \alpha} \times {\vec m}} \, e^{i {\vec \phi}_{pq} 
\cdot {\vec m}}}{(x + {\vec \bc} \cdot {\vec m} )^2}
\label{Vtwo}
\ee
while the spin self-coupling ${\tilde W}_{\vec \alpha}^{pq}$ is
\be
{\tilde W}_{\vec \alpha}^{pq} = \frac{1}{\m} \sum_{{\vec m} \neq {\vec 0}}
\frac{ \omega^{{\vec \alpha} \times {\vec m}} \, e^{i {\vec \phi}_{pq}
\cdot {\vec m}}}{({\vec \bc} \cdot {\vec m} )^2}
\label{self}
\ee 

If the above potentials were independent of the $U(\n)$ indices
$\vec \alpha$ and $p,q$, the sum over $U(\n)$ indices in the potential
energy expression (\ref{Vfin}) would simply be a $U(\n)$ trace 
and would give the $U(\n)$-invariant coupling between
the spins of particles $i$ and $j$ multiplying the standard Weierstrass
potential of the elliptic Calogero model. In the present case, however,
the above potential is spin-dependent and breaks $U(\n)$ invariance,
introducing a star-product twist in the indices $\vec \alpha$ 
and phase shifts ${\vec \phi}_p$ in the indices $p,q$.
Generically, the $U(\n)$ invariance of the original model is broken
down to an abelian $U(1)^k$, amounting to the transformation
\be
(S_i )_{\alpha \beta}^{pq} \to e^{i\varphi_p} \,
(S_i )_{\alpha \beta}^{pq} ~ e^{-i\varphi_q}
\ee
If ${\vec \phi}_p$ are equal for $k'$ values of $p$, the remaining 
symmetry $U(1)^{k'}$ is enhanced to $U( k' )$, corresponding 
to mixing the corresponding $p$-components.

The case $\omega=1$, ${\vec \phi}_p = 0$
reduces to the standard spin-elliptic Calogero-Moser model. The
case $\m = 1$ (and thus $\omega = 1$) reproduces the 
$U(\n)$-noninvariant model of the previous subsection. The general
case with $\omega \neq 1$ is a new classical integrable model
of the spin-Calogero type with a spin-dependent potential which
is a modular function of the two-body distance $x_{ij}$. 

The sums appearing in (\ref{Vtwo}) and (\ref{self})
could in principle have ambiguities due to the logarithmic divergence
of the summation over the radial coordinate on the complex plane.
This is, indeed, the case for the standard Weierstrass function and
a specific prescription is needed to regularize it. Different prescriptions
lead to different additive constants in the result. In our case,
however, the presence of the extra phases renders the sums convergent
and there is no regularization ambiguity.

The potentials can be expressed in terms of theta-functions.
$W_{\vec \alpha}^{pq} (x)$ is a modular function 
on the complex torus $(\ba ,\bb )$ with quasiperiodicity
\begin{eqnarray}
W_{\vec \alpha}^{pq} (x+\ba ) &=& e^{-{i\phi_{pq} + \frac{2\pi \mu}
{m} \alpha_2}} ~ W_{\vec \alpha}^{pq} (x) \nonumber \\
W_{\vec \alpha}^{pq} (x+\bb ) &=& e^{-{i\theta_{pq} - \frac{2\pi \mu}
{m} \alpha_1}} ~ W_{\vec \alpha}^{pq} (x) 
\end{eqnarray}
It has a double pole at $x=0$, with principal part
\be
W_{\vec \alpha}^{pq} (x) = \frac{1}{\m x^2} + O( x^0 )
\ee
and no other poles in each cell. These properties uniquely define
$W_{\vec \alpha}^{pq} (x)$ and allow for an expression in terms of 
theta-functions. We put
\be
W_{\vec \alpha}^{pq} (x) = A ~ \omega^{-i\frac{x}{\ba}} ~
e^{-{i\frac{x}{\ba}\phi_{pq}}} ~ \frac{
\tha \left( \frac{\pi}{\ba} (x-Q_1) \right)
\tha \left( \frac{\pi}{\ba} (x-Q_2)\right)}
{\tha \left( \frac{\pi}{\ba} x \right)^2}
\label{VAQQ}
\ee
where $Q_{1,2}$ are the as yet unknown zeros of $W_{\vec \alpha}^{pq} (x)$
and the theta-functions appearing above have complex period
$T = \bb /\ba$. This has the right quasiperiodicity under
$x \to x+\ba$. In order to also have the right quasiperiodicity under
$x \to x+\bb$, $Q_{1,2}$ must satisfy
\be
Q_1 + Q_2 = \frac{{\vec \phi_{ab}} \times {\vec \bc}}{2\pi}
+ \frac{\mu}{\m} {\vec \alpha} \cdot {\vec \bc}
\label{QQ}
\ee
and to have the right behavior around $x=0$ we must further have
\be
\frac{ \tha' \left( \frac{\pi}{\ba} Q_1 \right) }
{ \tha \left( \frac{\pi}{\ba} Q_1 \right) } +
\frac{ \tha' \left( \frac{\pi}{\ba} Q_2 \right) }
{ \tha \left( \frac{\pi}{\ba} Q_2 \right) }
= -i\frac{\phi_{ab}}{\pi} -2i \frac{\mu}{\m} \alpha_1
\label{QQp}
\ee
\be
A = \frac{\pi^2 \, \tha' (0)^2}{\m \, \ba^2 \,
\tha \left( \frac{\pi Q_1}{\ba} \right)
\tha \left( \frac{\pi Q_2}{\ba} \right)} 
\label{AQQ} 
\ee
The equations (\ref{QQ}) and (\ref{QQp}) above determine
$Q_1$ and $Q_2$, while (\ref{AQQ}) in turn determines $A$. 
The self-coupling ${\tilde W}_{\vec \alpha}^{pq}$ can then
be extracted from $W_{\vec \alpha}^{pq} (x)$ as
\be
{\tilde W}_{\vec \alpha}^{pq} = \lim_{x \to 0} 
\left( W_{\vec \alpha}^{pq} (x) - \frac{1}{\m\, x^2} \right)
\ee

The sums appearing in (\ref{Vtwo}) and (\ref{self}) are in general
convergent, due to the presence of the phases.
For $\omega =1$, however, the phases are absent and terms
with $p=q$ have an additive ambiguity due to the need for 
regularization for the expression (\ref{Vtwo}).
In the theta-function expression this manifests in the fact that
the equations for $Q_{1,2}$ (\ref{QQ},\ref{QQp})
are satisfied for {\it any} $Q_1 = -Q_2$.
By applying the addition formula
\be
\tha (x+Q) \tha (x-Q) \thh (0)^2 =  \tha (x)^2 \thh (Q)^2
- \thh (x)^2 \tha (Q)^2
\ee
this is seen indeed to amount to an arbitrary additive constant 
to the expression for $W^{pp} (x)$. The same holds for terms
$p,q$ for which ${\vec \phi}_{pq} = 0$. Such arbitrariness,
however, corresponds to trivial redefinitions of the model
by addition of constants of motion, as explained in the previous
subsection.

In conclusion, we identified an integrable generalization of the 
elliptic spin model which breaks the spin $U(n)$ invariance
and promotes the potential to a modular function introducing
noncommutative spin twists.

\section{Epilogue}
This concludes our promenade in Calogero land. We have seen and touched
many aspects of these models, but have by no means exhausted them.
There are various other issues that have not been visited. A few of
these are listed below.

There are systems with nearest-neighbor interactions that can also be obtained
and solved in the exchange operator formalism \cite{EFGR}. Such systems were
not analyzed here. We also mention the alternative operator approaches to
analyze such models used in \cite{GP}.

Spin models that can be solved with the techniques outlined
in this review include many examples not treated here. Apart from the
supersymmetric models mentioned earlier, we also list the models treated
in \cite{FGGRZ}.

The continuous limit of a system of Calogero particles can be described
in the collective field theory formulation \cite{AJL}. The system exhibits
interesting soliton and wave solutions \cite{WS,ABJ}, whose chiral nature
and connection to Benjamin-Davis-Acrivos-Ono solitons is an intresting
issue \cite{AW}.

Finally, there are certainly lots more topics related to the Calogero model,
and related references, to which this report has not made justice. The most
egregious omissions will hopefully be rectified in the next revision.


\begin{thebibliography}{99}

\bibitem{Calo} 
F.~Calogero, {\it Jour. of Math. Phys.}
{\bf 10}, 2191 and 2197 (1969); {\bf 12}, 419 (1971);
{\it Lett.\ Nuovo Cim.}  {\bf 13}, 411 (1975).
%%CITATION = NCLTA,13,411;%%

\bibitem{Suth}
B.~Sutherland, \PR {\bf A4}, 2019 (1971); \PR {\bf A5}, 1372 (1972);
\PRL {\bf 34}, 1083 (1975).

\bibitem{Mos}
J.~Moser, {\it Adv. Math.} {\bf 16}, 1 (1975).

\bibitem{KKS}
D.~Kazhdan, B.~Kostant and S.~Sternberg, {\it Comm. Pure Appl. Math.}
{\bf 31}, 481 (1978).

\bibitem{OP}
M.~A.~Olshanetskii and A.~M.~Perelomov, {\it Phys. Rep.} {\bf 71},
314 (1981); {\bf 94}, 6 (1983).

\bibitem{GHW} J.~Gibbons and T.~Hermsen, {\it Physica} {\bf D11} (1984) 337;
S.~Wojciechowski, \PL {\bf A111} (1985) 101.

\bibitem{PolA}
A.~P.~Polychronakos, \NP {\bf B324}, 597 (1989).
%%CITATION = NUPHA,B324,597;%%

\bibitem{Hal} F.~D.~M.~Haldane, \PRL {\bf 60}, 635 (1988) and
{\bf 66}, 1529 (1991).

\bibitem{Sha}
B.~S.~Shastry,\PRL {\bf 60}, 639 (1988) and {\bf 69}, 164 (1992).

\bibitem{Ino}
V.~I.~Inozemtsev, {\it J. Stat. Phys} {\bf 59}, 1143 (1989).

\bibitem{PolEX} 
A.~P.~Polychronakos, 
\PRL {\bf 69}, 703 (1992) [arXiv:hep-th/9202057].
%%CITATION = HEP-TH 9202057;%%

\bibitem{Rui}
S.~N.~M.~Ruijsenaars,
{\it Commun.\ Math.\ Phys.}  {\bf 110}, 191 (1987).
%%CITATION = CMPHA,110,191;%%

\bibitem{LesHouches}
A.~P.~Polychronakos,
``Generalized statistics in one dimension,''
Les Houches 1998 lectures [arXiv:hep-th/9902157].
%%CITATION = HEP-TH 9902157;%%

\bibitem{HP} E.~D' Hoker and D.~H.~Phong, \NP {\bf B530}, 537 
and 611 (1998).

\bibitem{FZ}
D.~B.~Fairlie and C.~K.~Zachos,
{\it Phys.\ Lett.} {\bf B224}, 101 (1989).
%%CITATION = PHLTA,B224,101;%%

\bibitem{LM}
J.M.~Leinaas and J.~Myrheim, \PR {\bf B37}, 9286 (1988).

\bibitem{Isa}
S.~B.~Isakov, \MPL {\bf B8}, 319 (1994) and \IJMP {\bf A9}, 2563 (1994).

\bibitem{Gau}
M.~Gaudin, Saclay preprint SPht-92-158.

\bibitem{MPCF}
J.A.~Minahan and A.~P.~Polychronakos, \PR {\bf B50}, 4236 (1994)
[arXiv:hep-th/9404192].
%%CITATION = HEP-TH 9404192;%%

\bibitem{LPS}
F.~Lesage, V.~Pasquier and D.~Serban, \NP {\bf B435}, 585 (1995).

\bibitem{Ha}
Z.N.C.~Ha, \PRL {\bf 73}, 1574 (1994). Erratum, ibid. {\bf 74}, 620 (1995).

\bibitem{GMM}
A.~P.~Polychronakos,
{\it Phys.\ Lett.} B {\bf B266}, 29 (1991).
%%CITATION = PHLTA,B266,29;%%

\bibitem{APQH}
A.~P.~Polychronakos,
{\it JHEP} {\bf 0104}, 011 (2001)
[arXiv:hep-th/0103013];
%%CITATION = HEP-TH 0103013;%%
{\bf 0106}, 070 (2001)
[arXiv:hep-th/0106011].
%%CITATION = HEP-TH 0106011;%%

\bibitem{HeRa}
S.~Hellerman and M.~Van Raamsdonk,
{\it JHEP} {\bf 0110}, 039 (2001)
[arXiv:hep-th/0103179].
%%CITATION = HEP-TH 0103179;%%

\bibitem{KaSa}
D.~Karabali and B.~Sakita,
{\it Phys.\ Rev.} {\bf B64}, 245316 (2001)
[arXiv:hep-th/0106016];
%%CITATION = HEP-TH 0106016;%%
{\bf B65}, 075304 (2002)
[arXiv:hep-th/0107168].
%%CITATION = HEP-TH 0107168;%%

\bibitem{MoP}
B.~Morariu and A.~P.~Polychronakos,
{\it JHEP} {\bf 0107}, 006 (2001)
[arXiv:hep-th/0106072];
%%CITATION = HEP-TH 0106072;%%
{\it Phys.\ Rev.} {\bf D72}, 125002 (2005)
[arXiv:hep-th/0510034].
%%CITATION = HEP-TH 0510034;%%

\bibitem{HKKCR}
T.~H.~Hansson, J.~Kailasvuori and A.~Karlhede,
\PR {\bf B68}, 035327 (2003)
[cond-mat/0109413];
A.~Cappelli and M.~Riccardi,
{\it J.\ Stat.\ Mech.}  {\bf 0505}, P001 (2005)
[arXiv:hep-th/0410151].
%%CITATION = HEP-TH 0410151;%%

\bibitem{IP}
V.~I.~Inozemtsev,
Phys.\ Scripta {\bf 39}, 289 (1989).
%%CITATION = PHSTB,39,289;%%
A.~P.~Polychronakos, \PL {\bf B277}, 102 (1992)
[arXiv:hep-th/9110064].
%%CITATION = HEP-TH 9110064;%%

\bibitem{PolC}
A.~P.~Polychronakos, \PL {\bf B408}, 117 (1997)
[arXiv:hep-th/9705047].
%%CITATION = HEP-TH 9705047;%%

\bibitem{NAM}
A physicist's exposition of relevant material is in Y.~Nambu,
``From $SU(3)$ to Gravity," eds. E.~Gotsman and G.~Tauber;
\PR {\bf D26} (1992) 2875, and references therein.

\bibitem{GN} A.~Gorsky and N.~Nekrasov, \NP {\bf B414}, 213 (1994).

\bibitem{MPYM} J.A.~Minahan and A.~P.~Polychronakos, \PL {\bf B326},
288 (1994) [arXiv:hep-th/9309044].
%%CITATION = HEP-TH 9309044;%%

\bibitem{Meh} H.~Mehta, {\it Random Matrices} (Academic Press,
New York, 1991).

\bibitem{FGP} J.~Fernandez Nunez, W.~Garcia Fuertes and A.M.~Perelomov,
arXiv:math-ph/0604062.

\bibitem{AAAP} A.~Agarwal and A.~P.~Polychronakos, {\it JHEP} (in print)
[arXiv:hep-th/0602049].
%%CITATION = HEP-TH 0602049;%%

\bibitem{Park} J.-H.~Park, \PL {\bf A307}, 183 (2003)
[arXiv:hep-th/0203017].

\bibitem{neg} M.A.~Olshanetskii and A.M.~Perelomov, {\it Invent.\ Math.}
{\bf 37}, 93 (1976);
L.~Feher and B.G.~Pusztai, \NP {\bf B734}, 304 (2006) [arXiv:math-ph/0507062];
arXiv:math-ph/0604073.

\bibitem{BL} J.~Blom and E.~Langmann, \PL {\bf B429}, 336 (1988).

\bibitem{PolR} A.~P.~Polychronakos, 
{\it Nucl.\ Phys.} {\bf B543}, 485 (1999)
[arXiv:hep-th/9810211].
%%CITATION = HEP-TH 9810211;%%

\bibitem{AP} 
A.~P.~Polychronakos, 
{\it Nucl.\ Phys.} {\bf B546}, 495 (1999)
[arXiv:hep-th/9806189].
%%CITATION = HEP-TH 9806189;%%.

\bibitem{Dunk} C.F.~Dunkl, {\it Trans. Amer. Math. Soc.} {\bf 311},
167 (1989).

\bibitem{BHV} L.~Brink, T.H.~Hansson and M.~Vassiliev, \PL {\bf B286},
109 (1992).

\bibitem{LV}
L.~Lapointe and L.~Vinet, \CMP {\bf 178}, 425 (1996).

\bibitem{MPSC} J.A.~Minahan and A.~P.~Polychronakos, \PL {\bf B302},
265 (1993)
[arXiv:hep-th/9206046].
%%CITATION = HEP-TH 9206046;%%

\bibitem{HH} Z.N.C.~Ha and F.D.M.~Haldane, \PR {\bf B46}, 9359 (1992).

\bibitem{Kaw} N.~Kawakami, \PR {\bf B46}, 1005 and 3191 (1992).

\bibitem{HW} K.~Hikami and M.~Wadati, \PL {\bf A173}, 263 (1993).

\bibitem{Hetc}
F.D.M.~Haldane, Z.N.C.~Ha, J.C.~Talstra, D.~Bernard and V.~Pasquier,
\PRL {\bf 69}, 2021 (1992).

\bibitem{BGHP}
D.~Bernard, M.~Gaudin, F.D.M.~Haldane and V.~Pasquier, {\it J. Phys}
{\bf A26}, 5219 (1993).

\bibitem{Cher}
I.~Cherednik, RIMS-1144 (1997) and references therein.

\bibitem{AJ}
J.~Avan and A.~Jevicki, \NP {\bf B469}, 287 (1996) and
{\bf B486}, 650 (1997).

\bibitem{PolD} A.~P.~Polychronakos, \NP {\bf B419}, 553 (1994)
[arXiv:hep-th/9310095].
%%CITATION = HEP-TH 9310095;%%

\bibitem{SS} B.~Sutherland and B.S.~Shastry, \PRL {\bf 70}, 4029 (1993).

\bibitem{FM}
M.~Fowler and J.A.~Minahan, \PRL {\bf 70}, 2325 (1993).

\bibitem{PolE} A.~P.~Polychronakos, \PRL {\bf 70}, 2329 (1993)
[arXiv:hep-th/9210109].
%%CITATION = HEP-TH 9210109;%%

\bibitem{Fra} H.~Frahm, {\it J. Phys} {\bf A26}, L473 (1993).

\bibitem{MX}
P.~Mathieu and Y.~Xudous,
{\it J.\ Phys.} {\bf A34}, 4197 (2001)
[arXiv:hep-th/0008036].
%%CITATION = HEP-TH 0008036;%%

\bibitem{KwK} N.~Kawakami and Y.~Kuramoto, 
\PR {\bf B50}, 4664 (1994).

\bibitem{KtK}
Y.~Kato and Y.~Kuramoto, 
\PRL {\bf 74}, 1222 (1995)
[cond-mat/9409031].

\bibitem{HBUW}
K.~Hikami and B.~Basu-Mallick,
\NP {\bf B566} [PM], 511 (2000)
[math-ph/9904033];
B.~Basu-Mallick, H.~Ujino and M.~Wadati,
{\it J.\ Phys.\ Soc.\ Jap.} {\bf 68}, 3219 (1999)
[arXiv:hep-th/9904167].
%%CITATION = HEP-TH 9904167;%%

\bibitem{FK} T.~Fukui and N.~Kawakami, YITP-95-17 (1996) and \NP
{\bf B483}, 663 (1997).

\bibitem{HG} F.D.M.~Haldane, \PL {\bf 93A}, 464 (1983) and 
\PRL {\bf 50}, 1153 (1983).

\bibitem{Weyl}
H.~Weyl, {\it Z. Phys.} {\bf 46} 1 (1927).

\bibitem{EFGR}
A.~Enciso, F.~Finkel, A.~Gonzalez-Lopez and M.~A.~Rodriguez,
{\it Phys.\ Lett.} {\bf B605}, 214 (2005)
[arXiv:hep-th/0407274];
%%CITATION = HEP-TH 0407274;%%
arXiv:nlin.si/0604073.
%%CITATION = NLIN-SI 0604073;%%

\bibitem{GP}
N.~Gurappa and P.~K.~Panigrahi, \PR {\bf B59}, R2490 (1999)
[cond-mat/9710035];
M.~Ezung, N.~Gurappa, A.~Khare and P.~K.~Panigrahi,
\PR {\bf B71}, 125121 (2005)
[cond-mat/0007005].
%%CITATION = COND-MAT 0007005;%%

\bibitem{FGGRZ}
F.~Finkel, D.~Gomez-Ullate, A.~Gonzalez-Lopez, M.~A.~Rodriguez and R.~Zhdanov,
{\it Commun.\ Math.\ Phys.}  {\bf 221}, 477 (2001)
[arXiv:hep-th/0102039];
%%CITATION = HEP-TH 0102039;%%
{\it Nucl.\ Phys.} {\bf B613}, 472 (2001)
[arXiv:hep-th/0103190].
%%CITATION = HEP-TH 0103190;%%

\bibitem{AJL}
I.~Andric, A.~Jevicki and H.~Levine,
{\it Nucl.\ Phys.} {\bf B215}, 307 (1983).
%%CITATION = NUPHA,B215,307;%%

\bibitem{WS}
A.~P.~Polychronakos,
{\it Phys.\ Rev.\ Lett.} {\bf 74}, 5153 (1995)
[arXiv:hep-th/9411054].
%%CITATION = HEP-TH 9411054;%%

\bibitem{ABJ}
I.~Andric, V.~Bardek and L.~Jonke,
{\it Phys.\ Lett.} {\bf B357}, 374 (1995)
[arXiv:hep-th/9411136].
%%CITATION = HEP-TH 9411136;%%

\bibitem{AW}
A.~G.~Abanov and P.~Wiegmann,
\PRL {\bf 95}, 076502 (1995) [cond-mat/0504041].





\end{thebibliography}
\end{document}